\documentclass[preprint,12pt,number]{elsarticle}

\usepackage{amsmath,amssymb}       
\usepackage{graphicx}              
\usepackage{booktabs}              
\usepackage{tabularx}              
\usepackage{longtable}             
\usepackage{multirow}              
\usepackage{makecell}              
\usepackage{xcolor}                
\usepackage{siunitx}               
\usepackage{microtype}             
\usepackage{setspace}              
\usepackage{float}                 
\usepackage{placeins}              
\usepackage{caption}
\usepackage{subcaption}            
\usepackage{listings}
\usepackage{etoolbox}
\usepackage[hidelinks]{hyperref}   

\lstdefinelanguage{json}{
  showstringspaces=false,
  literate=
    *{:}{{{\color{black}:}}}{1}
     {,}{{{\color{black},}}}{1}
     {\{}{{{\color{gray}\{}}}{1}
     {\}}{{{\color{gray}\}}}}{1}
     {[}{{{\color{gray}[}}}{1}
     {]}{{{\color{gray}]}}}{1},
  string=[s]{"}{"},
  stringstyle=\color{teal!70!black},
  comment=[l]{//},
  commentstyle=\color{gray!80}\itshape,
}

\lstdefinestyle{jsonstyle}{
  language=json,
  basicstyle=\ttfamily\footnotesize,
  numbers=none,
  breaklines=true,
  frame=single,
  framerule=0.4pt,
  rulecolor=\color{gray!40},
  backgroundcolor=\color{gray!5},
}
\AtBeginEnvironment{table}{\footnotesize}
\AtBeginEnvironment{longtable}{\footnotesize}
\renewcommand{\arraystretch}{1.15}
\biboptions{sort&compress}

\begin{document}

\begin{frontmatter}

\title{LoRIS: LoRaWAN-based IoT Platform for Sustainability Monitoring in Hotels}

\author[aff1]{Yash Pandey}
\author[aff1]{Angus Gray}
\author[aff1]{Reza Serati}
\author[aff2]{Oscar Zhu}
\author[aff4]{Emil Juvan}
\author[aff3]{Anna Zinn}
\author[aff3,aff5]{Danyelle Greene}
\author[aff3]{Qingqing Chen}
\author[aff3]{Sarah MacInnes}
\author[aff1]{Siamak Layeghy}
\author[aff3,aff4]{Sara Dolnicar}
\author[aff1]{Marius Portmann}


\affiliation[aff1]{
  organization={School of Electrical Engineering and Computer Science,
                The University of Queensland},
  city={Brisbane},
  state={QLD},
  country={Australia}
}

\affiliation[aff2]{
  organization={School of Management, Zhejiang University},
  city={Hangzhou},
  country={China}
}

\affiliation[aff3]{
  organization={UQ Business School,
                The University of Queensland},
  city={Brisbane},
  state={QLD},
  country={Australia}
}

\affiliation[aff4]{
  organization={The University of Primorska},
  city={Koper},
  country={Slovenia}
}

\affiliation[aff5]{
  organization={School of Psychology, Murdoch University},
  city={Perth},
  state={WA},
  country={Australia}
}

\begin{abstract}
The hospitality sector is a major source of global greenhouse gas emissions, water stress, and waste generation, yet sustainability reporting in hotels remains constrained by coarse, manually collected operational data. We present LoRIS (LoRaWAN-based IoT platform for sustainability monitoring in hotels), a LoRaWAN-based sensing system that delivers high-resolution measurements of resource consumption, environmental conditions, and guest behaviour across geographically distributed hotel properties. The architecture follows the canonical LoRaWAN reference model and is built for the operational realities of hospitality deployments: restrictive hotel IT policies, guest privacy expectations, rapid and reversible installation, and multi-year battery operation. Privacy-by-design guides modality selection and deployment zoning, and end-to-end encryption protects data from sensor to dashboard. This system has been running since February 2022 and currently spans 850 sensors of 19 types across 21 sites in Australia and Slovenia, covering both the AU915 and EU868 regulatory regions. The platform has generated over 202 million sensor records and ingests approximately 245,000 uplink messages per day on managed serverless infrastructure. Our system has been successfully used for seven field studies spanning food waste, energy consumption, and water consumption, including controlled intervention experiments that measure environmental outcomes and guest satisfaction in parallel. This system shows that LoRaWAN sensing can be deployed at scale in operational hotels without compromising guest experience or privacy.
\end{abstract}


\end{frontmatter}




\section{Introduction}
\label{sec:intro}

Hotels sit at a difficult intersection of around-the-clock service expectations and increasing pressure to reduce environmental impact. Tourism matters to the climate: the sector produced 5.2 gigatonnes of carbon dioxide equivalent (GtCO$_2$e) in 2019, or 8.8\% of global greenhouse gas emissions. Emissions grew because efficiency gains of 0.3\% per year were outweighed by demand growth of 3.8\% per year \cite{sun_drivers_2024}. Transport dominates this footprint, but shopping and food supply chains also contribute substantially \cite{lenzen_carbon_2018}. Holding warming to 1.5--2.0\textdegree C requires the sector to halve emissions by 2030 and reach net-zero by mid-century, which depends on comprehensive emission inventories and clear decarbonisation responsibilities for each stakeholder \cite{gossling_review_2023}. Accommodation is one of the few parts of that footprint a single operator can measure and act on directly.

Hotels are particularly relevant because they run continuously, hold indoor conditions within a narrow thermal comfort band, and cannot adopt sustainability measures that reduce guest satisfaction. Hotel resource consumption varies widely with outdoor temperature, service level, and room occupancy, which complicates fair comparisons and makes coarse annual reporting inadequate for actionable operations \cite{bohdanowicz_determinants_2007}. Building energy is a dominant lever, particularly heating, ventilation, and air-conditioning (HVAC), which accounts for 30--50\% of hotel energy use and motivates higher-granularity sensing and control that target the largest loads \cite{torres_heating_2020}. Accommodation services couple water and energy tightly through laundry, hot water, pools and spas, and kitchens, so tourist accommodation forms part of the water-energy nexus and operators must monitor and optimise both resources jointly rather than treat them as independent utilities \cite{becken_evidence_2017}.

Waste, particularly food waste, adds another major sustainability dimension. Hotel waste streams are large and complex \cite{juvan_waste_2023, pirani_solid_2014}, and effective reduction runs into operational barriers around segregation, measurement practices, incentives, and process design \cite{pirani_solid_2014}. The drivers of hotel food waste (buffet service design, forecasting and procurement, staff behaviour, guest expectations) are now well catalogued alongside mitigation practices \cite{juvan_drivers_2021, kasavan_drivers_2022}, yet measurement approaches remain fragmented across studies and properties, and most available data is self-reported by hotel management \cite{juvan_waste_2023}. Hotel buffet contexts generate substantial waste: field measurement at breakfast buffets links plate waste to guest mix and buffet layout \cite{juvan_biting_2018}, and continuous sensor-based measurement puts typical plate waste at 200--300 g per guest per day \cite{dolnicar_automatically_2023,juvan_importance_2025,zinn_not_2026}. Beyond hotels, food service accounts for 28\% of the 1.05 billion tonnes of food wasted globally in 2022, roughly 290 million tonnes, a figure the report describes as likely underestimated because sector-level measurement data remains sparse \cite{environment_food_2024}.

Reducing consumption in these domains increasingly relies on behaviour interventions: informational messages, social-norm framing, feedback, and other interventions that aim to shift how guests and staff use energy, water, and food \cite{dolnicar_designing_2020}. Evaluating whether these interventions work is itself a measurement problem. Social-science studies typically assess them through surveys that ask whether attitudes or behaviours changed, yet stated behaviour is a weak proxy for what people actually do \cite{juvan_measuring_2016}. Self-reported pro-environmental behaviour correlates only moderately with objectively observed behaviour, because social-desirability bias and inaccurate recollection inflate what people report \cite{kormos_validity_2014}. Even environmentally committed tourists show a persistent gap between environmental attitudes and holiday behaviour \cite{juvan_attitudebehaviour_2014}. Improving an intervention therefore requires measuring its effect directly, by observing consumption continuously and at its source rather than relying on what people report afterwards.

A core bottleneck across these domains is the lack of consistent,
high-resolution operational data for sustainability reporting and benchmarking in hospitality. Even among the largest global hotel groups, a substantial portion publish no sustainability reports, and reporting logics and structures change over time, which limits comparability and slows operational learning \cite{guix_changing_2025}. This motivates systems that instrument the building and operations, produce standardised metrics, and do so with minimal disruption and cost.

IoT sensing is a natural enabler because it connects distributed, heterogeneous assets (rooms, HVAC zones, water lines, kitchens, waste collection points) into a unified data layer. Foundational IoT surveys define this layer as the combination of identification, sensing and actuation, communication protocols, and distributed computation, rather than as sensors alone \cite{atzori_internet_2010,gubbi_internet_2013,al-fuqaha_internet_2015}. In smart buildings, this becomes an end-to-end systems problem: hundreds to thousands of sensing points must be ingested, cleaned, stored, visualised, and made available for analytics and decision support across comfort, energy, safety, and waste-management use cases \cite{bashir_reference_2022}. In hospitality, smart-hotel architectures similarly organise sensing, transport, preparation, processing, and application layers, or place fog nodes between hotel devices and cloud services to improve responsiveness,
local analytics, and scalability \cite{kansakar_fog-assisted_2018,hassan_shmis_2025}. These architectures establish the value of IoT for smart hospitality, but they provide limited evidence on long-running, privacy-sensitive, multi-property deployments whose primary objective is sustainability measurement rather than guest-facing automation.

For connectivity at hotel scale, Low Power Wide Area Networks (LPWANs) fill the gap left by short-range wireless networks and conventional cellular systems. They trade bandwidth and latency for long range, low device cost, and low energy operation, which matches sustainability sensing where measurements are small, periodic, and geographically distributed \cite{raza_low_2017,mekki_comparative_2019}. LoRaWAN is especially attractive for retrofit deployments because its long-range, low-power communication enables battery-operated sensors to be installed without new power or data cabling. End devices transmit through a small number of gateways, which can use existing Ethernet, Wi-Fi, or cellular connections to forward data to network and application servers. This architecture allows sensing infrastructure to be added incrementally to existing buildings with minimal disruption. The standard defines regional channel plans, device classes, frame counters, activation modes, security keys, and adaptive data rate mechanisms \cite{haxhibeqiri_survey_2018,noauthor_ts001-104_2023,noauthor_rp2-102_2020}. This model allows a hotel deployment to operate as a private sensing network with minimal dependence on property Wi-Fi, while still using IP-based backhaul from gateways to
managed cloud services.

At the same time, LoRaWAN is not a drop-in solution for dense sensing. Capacity depends on airtime, spreading factor assignment, gateway density, channel selection, duty-cycle restrictions, and interference \cite{bor_lora_2016,georgiou_low_2017,adelantado_understanding_2017}. Surveys of LoRa networking and scalable LoRaWAN identify link coordination, resource allocation, reliable transmission, downlink scarcity, ADR behaviour, and security as persistent research and deployment issues \cite{shanmuga_sundaram_survey_2020,jouhari_survey_2023}.
Energy models further show that reporting frequency, spreading factor, retransmissions, and receive-window behaviour determine battery lifetime, making protocol configuration an operational design parameter rather than a purely networking detail \cite{casals_modeling_2017,kufakunesu_survey_2020}.

Evidence from deployed LoRa and LoRaWAN systems shows that these constraints appear in operation rather than only in protocol analysis. CityWAN provides a useful upper bound for research-scale LoRa networking, reporting a citywide deployment with about 19.8k LoRa nodes, 100 gateways, 130 km$^2$ of coverage, and 12 smart-city applications, although it is a LoRa system rather than a clearly standard LoRaWAN deployment \cite{tong_citywide_2024}. Production LoRaWAN utility networks can be even larger: the Paphos smart-water network contains more than 30,000 LoRaWAN smart water meters and has been analysed using 12 months of operational data, but it is best understood as a production metering network studied after deployment rather than a research-built sensing platform \cite{lavdas_evaluating_2025,lavdas_performance_2025}. Between these extremes, A2A Smart City in Brescia has installed about 3000 LoRaWAN sensors and more than 500 publicly accessible devices across smart parking, smart metering, and smart-bin applications \cite{gaffurini_end--end_2024}. Long-running research-managed LoRaWAN deployments are closer to the operating conditions of our system but are typically smaller or narrower in scope. The University of Oulu smart campus deployed 331 LoRaWAN nodes, comprising 1655 sensing elements, for more than two years in an indoor campus environment; the study showed that packet loss arose not only from the radio link but also from the IP backbone, external interference, nonuniform transmission timing, and seasonal effects \cite{yasmin_lorawan_2020}. Southampton's smart-city deployment analysed 20 devices and more than 135,000 messages, with gateway locations constrained by network access, permissions, and accessibility rather than theoretical optimality \cite{basford_lorawan_2020}. A live multi-gateway deployment in Glasgow similarly showed that indoor and outdoor LoRaWAN performance can differ sharply, with packet loss affected by interference, channel choice, and backhaul behaviour \cite{harinda_performance_2022}. These systems motivate deployment designs that log network metadata as well as sensing values, tolerate missing or delayed packets, and treat gateway placement, backhaul, interference, and maintenance access as first-class systems constraints.

Applying this research to a hotel-wide sustainability data system remains challenging because hotels face practical constraints that are not fully represented in laboratory, smart-city, or campus testbeds. First, hotel IT systems are increasingly complex and exposed to cybersecurity risks. As a result, hotels often have strict policies on connecting third-party devices to their internal networks, controlling system access, and tracking changes and activity
\cite{wynn_it_2022,karadayi-usta_cybersecurity_2024}. Second, sensing intersects with guest privacy: privacy management (policy, assurance, and internal access control) shapes guest trust and willingness to accept information collection \cite{moon_hotel_2022}, and smart-building sensing creates privacy threats when sensor modalities, placement, and data handling are not explicitly privacy-aware
\cite{sakariyah_adewole_systematic_2025}. Third, segmented and legacy building networks constrain backhaul connectivity, so deployments must tolerate intermittent links and delayed on-site remediation \cite{chan_iot_2023,frei_building_nodate}. Fourth, battery longevity becomes an operational bottleneck at scale, and LoRaWAN device lifetime
depends on protocol choices and retransmissions \cite{casals_modeling_2017}. Finally, installation must be fast and non-destructive because rooms are revenue-generating spaces, yet field deployments in occupied buildings face mounting trade-offs:
removable attachment increases detachment risk, while stronger adhesives increase surface-damage risk \cite{frei_building_nodate}.

We address these constraints with LoRIS, our LoRaWAN-based sensing system for hotel-scale sustainability monitoring. Our design targets least-privilege data access, privacy-preserving choice of sensing modalities, robustness to intermittent backhaul, multi-year battery operation under realistic LoRaWAN settings, and rapid, reversible mounting compatible with housekeeping workflows. In this paper, we describe the platform architecture, document its operational performance across more than four years of continuous deployment (850 sensors at 21 sites in Australia and Slovenia, over 202 million records since February 2022), and demonstrate its successful use through seven field studies spanning food waste, energy consumption, and water consumption.

\section{System design and architecture}
\label{sec:architecture}

The system is built on managed cloud services and commercial off-the-shelf IoT components, configured for the operational and privacy constraints of hotel deployments. We designed the system to provide a scalable, multi-tenant, and privacy-preserving platform for sustainability data collection, processing, and visualisation across geographically distributed properties. Sensor data and derived sustainability metrics help hotel managers, sustainability officers, and researchers make data-driven decisions about resource consumption, environmental conditions, and guest behaviour at site level.

\begin{figure}[H]
  \centering
  \includegraphics[width=\columnwidth]{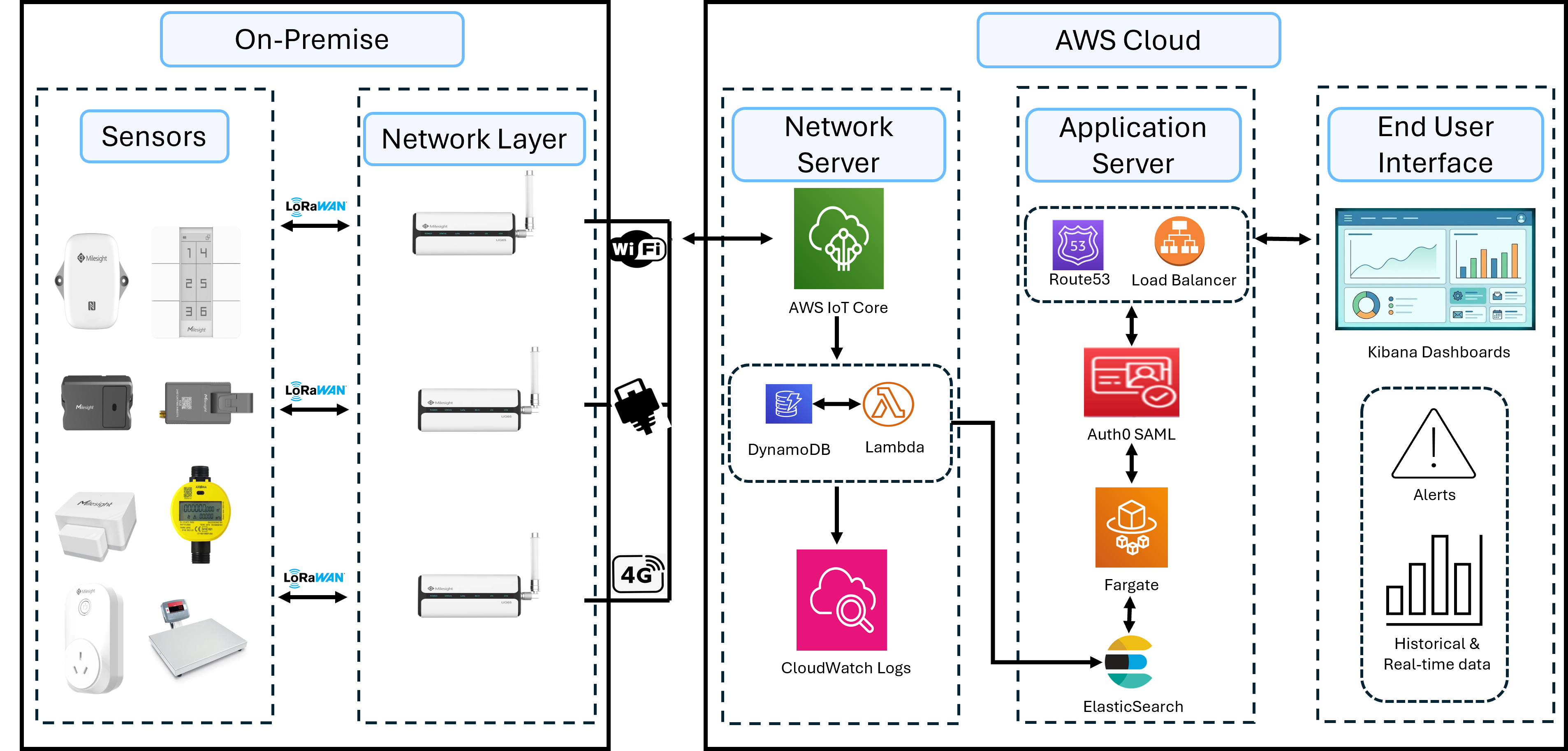}
  \caption{LoRIS system architecture and data flow across the LoRaWAN reference components.}
  \label{fig:architecture}
\end{figure}

Fig.~\ref{fig:architecture} illustrates the architecture of our system. It comprises five components: sensors, network layer, the network server, application server, and end-user interface. Security and privacy controls span across all five components. The system has been running continuously since February 2022 and currently has 850 sensors across 21 sites in Australia and Slovenia, with over 202 million sensor records accumulated to April 2026.

\subsection{Sensors}
\label{sec:sensors}
Sensors collect data from guest rooms, common areas, kitchens, service corridors, and waste collection points. We built the deployment around commercial LoRaWAN-enabled sensors from four manufacturers (Milesight, Netvox, Axioma, Elsys), supplemented by one custom-built sensor (floor scales for food waste measurement), comprising 19 sensor types in total. Power source and LoRaWAN device class follow from the installation context rather than the sensor type. Battery-powered sensors use replaceable 3.6~V lithium-thionyl-chloride (ER14505) cells and operate in Class A, opening two short receive windows after each uplink. Two variations exist within this group: the Axioma W1 water meters carry cells sealed for the service life of the meter, and the Milesight Smart CT10x current transformers need no battery at all, drawing power from the conductor they measure. Mains-powered sensors operate in Class C, holding the receive window open between uplinks so that configuration downlinks take effect immediately. This group covers the AM319 air quality sensors, the WS523 and R809A smart plugs, and the UC300 bridges in the 26 platform scales. The Milesight VS121 people counters are the exception: they draw mains power but operate in Class A. No sensor in the deployment uses Class B. Table~\ref{tab:sensors} lists detailed information for all the sensors, categorised in the following three types: environmental data sensors, resource consumption data sensors, occupancy and event data sensors, discussed in more detail below.

\begin{table}[H]
\centering
\footnotesize
\setlength{\tabcolsep}{4pt}
\renewcommand{\arraystretch}{1.15}
\caption{Sensor details, including model name, manufacturer, sensor modalities, number of units deployed, and number of deployment sites.}
\label{tab:sensors}
\begin{tabular}{p{2.2cm} p{1.7cm} p{5.3cm} >{\centering\arraybackslash}p{1.5cm} >{\centering\arraybackslash}p{1.1cm}}
\toprule
\textbf{Sensor model} & \textbf{Manu-}\newline\textbf{facturer} & \textbf{Sensor modality} & \textbf{Deployed sensors} & \textbf{Sites} \\
\midrule
\multicolumn{5}{l}{\textit{Environmental data sensors}} \\
\midrule
AM319 & Milesight & Temperature, relative humidity, light intensity, air quality (CO$_2$, TVOC, PM2.5/PM10) & 16 & 4 \\
EM300-TH & Milesight & Temperature, relative humidity & 395 & 21 \\
ERS-Sound & Elsys & Sound intensity (dB), temperature, relative humidity, light \& motion & 9 & 3 \\
\midrule
\multicolumn{5}{l}{\textit{Resource consumption data sensors}} \\
\midrule
W1 & Axioma & Water flow rate (cubic metres/hour) and volume (cubic metres) & 6 & 4 \\
Smart CT10x & Milesight & Electrical current (RMS and accumulated) & 28 & 3 \\
R718N1 & Netvox & Electrical current (single phase, external CT) & 1 & 1 \\
WS523 & Milesight & Voltage, electrical current \& energy/power & 13 & 6 \\
R809A & Netvox & Voltage, electrical current \& energy/power & 106 & 6 \\
EM310-UDL & Milesight & Distance (bin fill level) & 3 & 1 \\
EM400-X (series) & Milesight & Distance (bin fill level) & 8 & 4 \\
WS201 & Milesight & Distance (bin fill level) & 1 & 1 \\
Platform scale\textsuperscript{a} & Custom & Weight (food/plate waste) & 26 & 9 \\

\midrule
\multicolumn{5}{l}{\textit{Occupancy and event data sensors}} \\
\midrule
VS121 & Milesight & People counting, occupancy & 8 & 7 \\
WS101 & Milesight & Button press (event trigger, single button) & 162 & 7 \\
WS156 & Milesight & Button press (event trigger, six buttons) & 34 & 3 \\
WS301 & Milesight & Door/window open/close state & 34 & 9 \\
\midrule
\multicolumn{3}{l}{\textbf{Total sensors}} & \textbf{850} & \\
\bottomrule
\multicolumn{5}{@{}p{13cm}@{}}{\footnotesize\textsuperscript{a}Industrial weighing scale integrated with an RS232-to-LoRaWAN bridge (Milesight UC300).} \\
\end{tabular}
\end{table}

\begin{itemize}
  \item \textit{Environmental data sensors}: indoor environmental data including temperature, relative humidity, air quality (CO$_2$, TVOC, PM2.5/PM10), light intensity, and sound level. The majority of the deployment (395 sensors) consists of Milesight EM300-TH temperature and humidity sensors, which transmit at one-minute intervals to support behavioural inference tasks such as shower event detection (transient humidity spikes resolved within 5--15 minutes).

  \item \textit{Resource consumption data sensors}: sensors monitoring electrical energy, water consumption, and waste-related parameters. Smart plugs (Milesight WS523, Netvox R809A), current meters (Milesight Smart CT10x, Netvox R718N1), and water meters transmit at 10--15 minute intervals. For food-waste monitoring, we built custom platform scales (26 units) by integrating an industrial weighing scale with an RS232-to-LoRaWAN bridge; these scales transmit on weight-change events. Milesight EM400-series distance sensors (ultrasonic EM400-UDL/MUD and laser time-of-flight EM400-TLD) measure fill level for general waste, plate/food waste, and recycling bins, reporting at 10-minute intervals.

  \item \textit{Occupancy and event data sensors}: sensors that detect human presence or discrete activity. Milesight VS121 people-counters (8 units) report aggregate occupancy in common areas, processing video frames on-device and transmitting only the resulting count. Event-driven sensors transmit only when triggered, including smart buttons (e.g., WS101) used as ground-truth labels in shower duration studies and contact sensors (e.g., WS301) for door and window state.
\end{itemize}

We selected sensors to suit hotel operating conditions, respecting guest privacy without disturbing housekeeping or kitchen operations. The selected sensors satisfy four requirements. First, privacy preservation through modality selection and deployment zoning (detailed in Section~\ref{sec:security}). Second, physical unobtrusiveness: we mount sensors using removable adhesive in under two minutes per room. Third, long-lasting autonomous operation: battery-powered sensors operate for more than one year between replacements. Fourth, zero integration with hotel IT infrastructure: all sensors communicate exclusively via LoRaWAN.

\subsection{Network layer}
\label{sec:network_layer}
The network layer carries data from sensors to the network server. Sensors transmit raw payloads to one or more site-local gateways using LoRaWAN. We use three Milesight gateway models selected by the installation environment of each site: the UG63 mini gateway (IP30 rated) at small sites and to fill indoor coverage blind spots, the UG65 semi-industrial gateway (IP65 rated) in standard indoor service areas, and the UG67 (IP67 rated) at outdoor and weather-exposed mounting positions. All three carry 8-channel Semtech SX1302 concentrators, support more than 2000 nodes, and run the LoRa Basics Station packet forwarder. The UG65 and UG67 run it on embedded Linux; the UG63 uses a microcontroller-based firmware stack. They present a uniform interface to the network server regardless of gateway model. The gateways support Ethernet and 4G cellular backhaul, with Wi-Fi additionally available on the UG65 and UG67. We choose Ethernet or Wi-Fi backhaul where site infrastructure allows, and fit gateways with an IoT SIM card as a 4G fallback where wired or Wi-Fi connectivity is unavailable or unreliable. Gateways forward received LoRaWAN frames via IP to the network server using the LoRaWAN Network Server (LNS) protocol over WebSocket Secure (WSS). The Configuration and Update Server (CUPS) protocol over HTTPS handles credential rotation and firmware updates.

Each gateway opens a persistent outbound WSS connection to the network server and holds it open for both uplink and downlink traffic. This means hotel firewalls do not need to permit any inbound connections: the network server pushes downlinks (Adaptive Data Rate (ADR) and other MAC-layer commands, plus application-layer messages) to the gateway over the same connection the gateway initiated. This pattern avoids integration with hotel IT infrastructure, a critical requirement given the restrictive IT policies of hospitality operators \cite{wynn_it_2022}.

We install gateways in service areas with access to mains power, mounting them at elevated positions (typically high on a wall or near the ceiling) to improve line-of-sight to sensors across guest room floors. The deployment spans two LoRaWAN regulatory regions: AU915 (Australia, no duty cycle restriction) and EU868 (Slovenia, 1\% duty cycle per sub-band). A routine EM300-TH uplink carries a 7-byte application payload, which LoRaWAN framing expands to a 20-byte PHY payload. At spreading factor 7 and 125 kHz bandwidth (SF7/BW125) each transmission has a time-on-air of $\sim$57~ms, giving a per-sensor duty cycle of $\sim$0.09\% at the one-minute reporting interval, an order of magnitude below the EU868 budget. Two factors add to this figure. Confirmed uplinks are enabled across the deployment (Section~\ref{sec:reliability}), so a sensor retransmits once when an acknowledgement does not arrive, and each acknowledgement consumes gateway airtime that the same duty cycle budget constrains. Time-on-air also grows sharply with spreading factor: the same payload occupies roughly 1.3~s at SF12, which no longer fits a one-minute interval under the EU868 budget. The one-minute interval is therefore feasible in Slovenia only while ADR holds a sensor at a low spreading factor, which is what we observe across the fleet (Section~\ref{sec:reliability}). We verified signal coverage at each site during commissioning to confirm adequate propagation throughout the property.

A practical limitation of this approach is that gateways operate as stateless packet forwarders. If the IP backhaul is interrupted, received LoRaWAN frames are dropped at the gateway until backhaul connectivity is restored. At one-minute reporting intervals, short outages cause little information loss. Indoor temperature and humidity vary over timescales of minutes to hours, so consecutive readings are strongly correlated and an isolated missing frame removes almost no unique information. Longer outages appear as discontinuities in the LoRaWAN uplink frame counter (FCnt): each sensor increments the counter on every transmission regardless of delivery, so the gap between the last counter value received before an outage and the first value received allows us to identify data gaps.

\subsection{Network server}
\label{sec:network_server}

The network server bridges the on-premise gateways and the application server. Each uplink message reaching the network server carries only a device identifier (DevEUI) and an encrypted hexadecimal payload, neither of which is directly usable for sustainability analytics. The network server turns this raw stream into structured, contextualised records through three sequential tasks: ingesting LoRaWAN protocol traffic, decoding per-sensor binary payloads, and enriching each record with deployment metadata. We implement these tasks on managed serverless AWS infrastructure, so the pipeline scales with sensor count without capacity planning.

AWS IoT Core for LoRaWAN serves as the network server, handling device activation (the join procedure through which a sensor and the network server establish a shared session), session key management, frame counter validation, multi-gateway deduplication, ADR, and downlink scheduling. LoRaWAN activation yields two AES-128 session keys: a network session key that protects message integrity between sensor and network server, and an application session key that encrypts the payload end to end. Because AWS IoT Core holds both keys, it decrypts application payloads on ingestion: the LoRaWAN encryption boundary ends at the managed network server rather than at a downstream application component. We accept this trust concentration as the cost of a managed service and scope access beyond this point through Identity and Access Management (IAM) policies (Section~\ref{sec:security}). Sensor frame events flow via AWS IoT Rules (a message-routing engine) to an AWS Lambda function (Node.js) that implements a sensor-type-specific decoder registry, currently containing 19 decoders. Each decoder, keyed by manufacturer and model identifier (e.g., \texttt{milesight/em300-th}, \texttt{netvox/r809a}), converts the raw hexadecimal payload into structured JSON. We preserve raw payloads alongside decoded values, which lets us reprocess historical data when vendor firmware updates change payload formats.

An AWS Lambda function then enriches each decoded record with deployment metadata retrieved from a DynamoDB table: site, building, room, position within the room (e.g., ``bathroom ceiling'', ``minibar'', ``bedside''), calibration parameters, and research project assignment. This enrichment produces a fully contextualised record ready for sustainability analytics. An AWS Lambda function forwards each enriched record to the application server for storage and visualisation (Section~\ref{sec:application_server}).

The pipeline currently processes approximately 245,000 uplinks per day. Amazon CloudWatch aggregates logs and operational metrics across the ingestion pipeline.

\subsection{Application server}
\label{sec:application_server}

The application server stores enriched records, hosts the visualisation backend, and mediates authenticated access for end users. Elastic Cloud (Elasticsearch) \cite{noauthor_elasticsearch_nodate} serves as the primary analytical store. Each indexed document carries sensor metadata, LoRaWAN network parameters (spreading factor, received signal strength indicator (RSSI), signal-to-noise ratio (SNR), frame counter, per-gateway reception details, etc.), and decoded sensor measurements with a UTC timestamp.

A containerised application on AWS Fargate, a serverless container runtime, running within a Virtual Private Cloud, mediates user access to the visualisation stack. Route 53 DNS and an Application Load Balancer route public traffic to the application, with the load balancer terminating Transport Layer Security (TLS) connections using certificates from AWS Certificate Manager. We store container images in Amazon Elastic Container Registry and deploy them via Continuous Integration and Continuous Delivery/Deployment (CI/CD) workflows from GitHub.

Auth0 \cite{noauthor_auth0_nodate} handles user authentication via Security Assertion Markup Language (SAML) federation. We enforce a role-based access control (RBAC) model that restricts data visibility by organisational scope: hotel managers access only their own property, researchers access cross-site data for specific projects, and administrators have system-wide access. We implemented a multi-tenant architecture in which each hotel is an independent logical entity, with access restricted to authorised users.

\subsection{End user interface}
\label{sec:end_user}

\begin{figure}[H]
  \centering
  \includegraphics[width=\columnwidth]{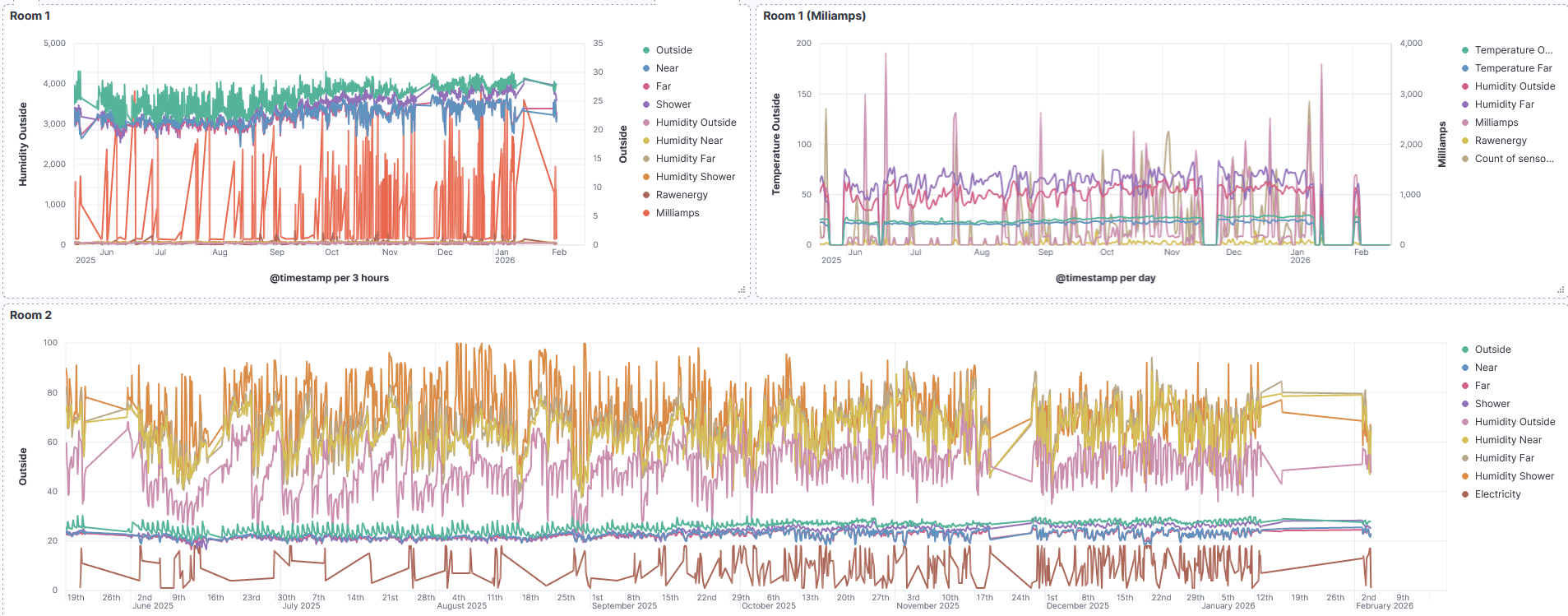}
  \caption{Example Kibana dashboard view from one of the partner hotels, showing co-located environmental and energy time-series for two instrumented guest rooms. Each panel overlays multiple sensor positions (outside, near, far, shower) and resource-consumption channels, allowing users to inspect long-term trends and short-term events at room level.}
  \label{fig:dashboard}
\end{figure}

End users interact with the system through Kibana dashboards \cite{noauthor_kibana_nodate} layered over the Elasticsearch store. The dashboards present site-level and room-level views of environmental conditions, resource consumption, and intervention outcomes, and each user configures their own real-time streams and historical-trend views (Fig.~\ref{fig:dashboard}). The system also generates automated email alerts on predefined thresholds (e.g., sustained temperature deviations, low sensor battery, frame counter discontinuities).

\subsection{Security and privacy}
\label{sec:security}

Two constraints shape our system's security and privacy design. Hospitality operators maintain restrictive IT policies for third-party devices \cite{wynn_it_2022, karadayi-usta_cybersecurity_2024}, so the platform cannot rely on hotel network trust or require firewall changes. Guest acceptance of in-room sensing depends on demonstrable privacy safeguards \cite{moon_hotel_2022}, so the platform must avoid collecting personally identifiable information by design rather than by policy. We describe the resulting controls in the order data flows through the system: privacy-preserving collection at the sensors, encrypted transit across the LoRaWAN link and backhaul, isolation within the cloud pipeline, and access control and governance at the point of use.

Privacy protection begins at the point of collection. We minimise sensitive data collection through sensor selection and deployment zoning: we classify the areas of each property by privacy sensitivity and restrict which sensor modalities each zone admits. Guest rooms, the most privacy-sensitive zone, receive only environmental sensors, smart power plugs, and event sensors (e.g., contact sensors, fridge open/close); no image-capable, audio-capable, or identity-linked sensors are present. Common areas such as dining halls admit people-counting sensors (Milesight VS121), which process video frames on-device through an embedded neural network and transmit only the resulting count. No image data leaves the sensor or is stored. Signage informs guests of environmental monitoring in instrumented rooms.

From sensor to network server, LoRaWAN AES-128 encryption secures the radio link. Device activation yields two session keys: the network session key (NwkSKey) ensures message integrity between sensor and network server, and the application session key (AppSKey) encrypts the payload end to end, so gateways forward ciphertext only and cannot read sensor data. We further treat the hotel network as untrusted: gateways open persistent outbound WSS connections to the network server and expose no inbound services, minimising the on-premise attack surface and removing the need for hotel firewall modifications. TLS 1.2+ protects the backhaul itself, with CUPS handling gateway credential rotation (Section~\ref{sec:network_layer}).

Within the cloud, the LoRaWAN encryption boundary ends at the network server: AWS IoT Core holds both session keys and decrypts application payloads on ingestion. We accept this trust concentration as the cost of a managed service and scope everything beyond it through Identity and Access Management (IAM) policies that enforce least-privilege access for each service component. TLS 1.2+ protects inter-service communication, and AWS-managed AES-256 encrypts data at rest in DynamoDB and Elasticsearch.

At the point of use, Auth0 SAML federation authenticates users, and the role-based access control model (Section~\ref{sec:application_server}) restricts data visibility by organisational scope, with each hotel isolated as an independent tenant. HTTPS/TLS secures all user sessions and API calls. Governance controls cover the residual risk that technical measures cannot remove: environmental time-series at one-minute resolution can in principle reveal occupancy patterns and daily routines, so we restrict raw-data access to authorised researchers under approved ethics protocols and pseudonymise room identifiers in published datasets.

\subsection{Data schema}
\label{sec:Data_schema}

Each record in the dataset corresponds to a single sensor uplink message and contains four categories of information: device and deployment metadata, LoRaWAN network metadata, gateway reception metadata, and decoded sensor measurements. Table~\ref{tab:schema} summarises the key fields with an example record as stored in Elasticsearch.

\begin{table}[H]
\centering
\scriptsize
\setlength{\tabcolsep}{5pt}
\renewcommand{\arraystretch}{1.05}
\caption{Example sensor uplink record as stored in Elasticsearch (a Milesight EM300-TH sensor in a guest-room shower at Countryside estate hotel, 19 May 2026), grouped by the four schema categories. Fields are stored under flattened dotted paths (\texttt{metadata.device.*}, \texttt{metadata.LoRaWAN.*}, \texttt{metadata.gateways.N.*}, \texttt{sensor.*}).}
\label{tab:schema}
\begin{tabular}{p{3.1cm} p{3.4cm} p{6cm}}
\toprule
\textbf{Field} & \textbf{Example value} & \textbf{Description} \\
\midrule
\multicolumn{3}{l}{\textit{Device metadata}} \\
\midrule
\texttt{DevEui} & \texttt{24e124136d088993} & Unique device identifier \\
\texttt{manufacturer}, \texttt{model} & \texttt{milesight}, \texttt{em300th} & Hardware type \\
\texttt{location}, \texttt{room} & \texttt{Countryside estate hotel}, \texttt{101} & Deployment site and room identifier \\
\texttt{position} & \texttt{shower} & Placement context within the room \\
\texttt{region} & \texttt{EU868} & LoRaWAN regional parameters \\
\texttt{organisation}, \texttt{project} & \texttt{slovenia}, \texttt{sdl} & Logical grouping for access control \\
\midrule
\multicolumn{3}{l}{\textit{LoRaWAN metadata}} \\
\midrule
\texttt{SpreadingFactor} & \texttt{7} & Spreading factor (SF7--SF12); indicates link conditions \\
\texttt{Bandwidth} & \texttt{125} & Channel bandwidth (kHz) \\
\texttt{DataRate} & \texttt{5} & LoRaWAN data rate index \\
\texttt{Frequency} & \texttt{868.3} & Channel centre frequency (MHz) \\
\texttt{FCnt} & \texttt{69581} & Frame counter (deduplication, replay detection) \\
\texttt{ADR} & \texttt{true} & Adaptive Data Rate enabled \\
\texttt{MType} & \texttt{ConfirmedDataUp} & LoRaWAN message type \\
\texttt{DevAddr} & \texttt{0032881d} & Session-scoped device address \\
\midrule
\multicolumn{3}{l}{\textit{Gateway reception}} \\
\midrule
\texttt{GatewayEui} & \texttt{24e124fffef5c590} & Receiving gateway identifier \\
\texttt{model}, \texttt{manufacturer} & \texttt{ug67}, \texttt{milesight} & Gateway hardware type \\
\texttt{location} & \texttt{Countryside estate hotel} & Gateway deployment site \\
\texttt{geo\_location} & \texttt{[xx.xxxx, yy.yyyy]} & Gateway coordinates (lon, lat) \\
\texttt{RSSI}, \texttt{SNR} & \texttt{-116}, \texttt{0.5} & Per-gateway reception quality \\
\midrule
\multicolumn{3}{l}{\textit{Sensor readings}} \\
\midrule
\texttt{sensor.temperature} & \texttt{21.91} & Decoded temperature (\textdegree C) \\
\texttt{sensor.humidity} & \texttt{46} & Decoded relative humidity (\%) \\
\texttt{payload} & \texttt{0367db0004685c} & Raw hexadecimal sensor payload (preserved for reprocessing) \\
\texttt{@timestamp} & \texttt{2026-05-19T02:18:37.000Z} & UTC timestamp of reception \\
\bottomrule
\end{tabular}
\end{table}

This schema preserves both decoded sensor values and the raw payload, which lets us reprocess historical records when device firmware updates change payload formats. Per-gateway reception metadata is stored as nested fields, supporting multi-gateway coverage analysis without inflating record counts. Dataset scale in Section~\ref{sec:data}, known data gaps in Section~\ref{sec:known_gaps}, and message delivery reliability in Section~\ref{sec:reliability}.

\section{System deployment}
\label{sec:implementation}

\subsection{Deployment sites}

We conducted a site survey at each site before installation. With the gateway mounted at its planned position (an elevated point in a service area with mains power, Section~\ref{sec:network_layer}), we walked the property with a Milesight FT101 field tester, transmitting test uplinks from every planned sensor location, including guest rooms, common areas, kitchens, and waste collection points. The tester reports the RSSI and SNR values of each transmission alongside a three-bar signal-strength indicator, and we included a location in the installation plan only after confirming adequate signal quality there. Where a location failed this check, we asked the hotel to substitute a comparable room, which preserved the experimental design of the study. Where the location itself was essential and could not be substituted, we installed an additional gateway to cover that area. This procedure keeps the fleet on the fastest LoRaWAN data rate: 97.59\% of deployed sensors operate at SF7, which minimises time-on-air and battery consumption.

\begin{figure}[H]
  \centering
  \includegraphics[width=\columnwidth]{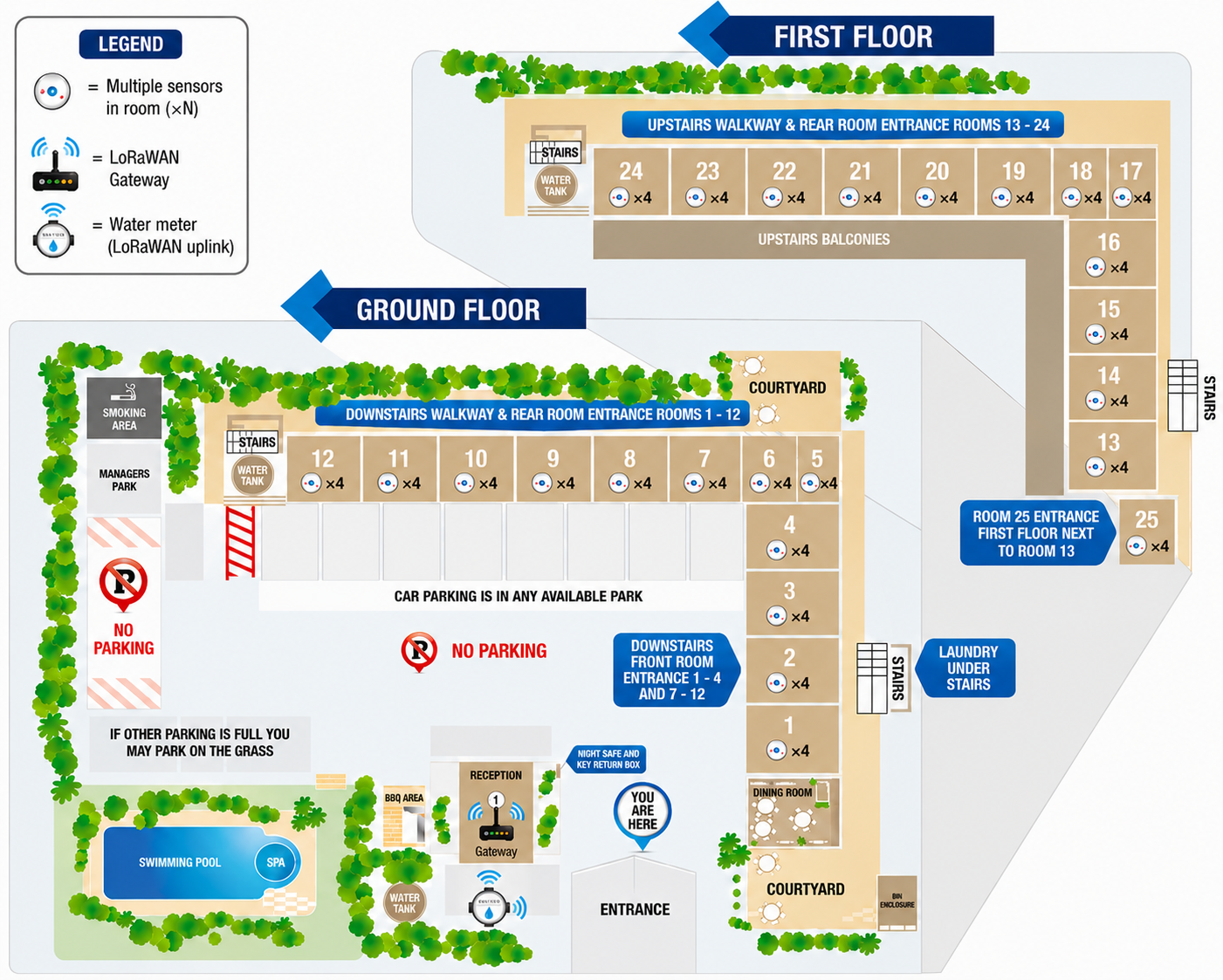}
  \caption{Example instrumented hotel floor plan showing room-level sensor deployments, a reception-based LoRaWAN gateway, and a LoRaWAN-enabled water meter near the ground-floor entrance. Each guest room is annotated with the number of deployed sensors.}
  \label{fig:floor_plan}
\end{figure}

We deployed this system across 21 sites in Australia (11 sites) and Slovenia (10 sites), covering both the AU915 and EU868 LoRaWAN regulatory regions and contrasting climate zones. The deployment comprises four property categories: 13 hotels (the largest category, spanning resort, city, and seaside properties in both countries), two student accommodation sites (one Australian university residential college and one Slovenian student hostel that also operates as a guest house in July and August), one budget motel in Australia, and five internal test and staging locations used for sensor commissioning, decoder development, and pre-deployment validation. Sensor counts per site range from a single sensor at sites instrumented for a specific study to 235 sensors at the largest student accommodation site, reflecting both property size and the scope of each engagement. We installed sensors based on the floor plan of each property and the experimental requirements of the studies it supported, typically prioritising guest rooms, common areas (dining halls, lobbies), kitchens, and waste collection points. Table~\ref{tab:deployment} lists per-site sensor counts and the sensor types deployed at each site. Fig.~\ref{fig:floor_plan} illustrates an example property-level deployment plan for one Australian motel site. The figure shows how sensors were distributed across guest rooms, with four sensors installed per room, together with the reception-based LoRaWAN gateway and a LoRaWAN-enabled water meter near the ground-floor entrance. This example demonstrates how deployments were adapted to the physical layout of each property while maintaining room-level coverage.

\begin{figure}[t]
  \centering
  \begin{subfigure}[t]{0.24\textwidth}
    \includegraphics[width=\linewidth]{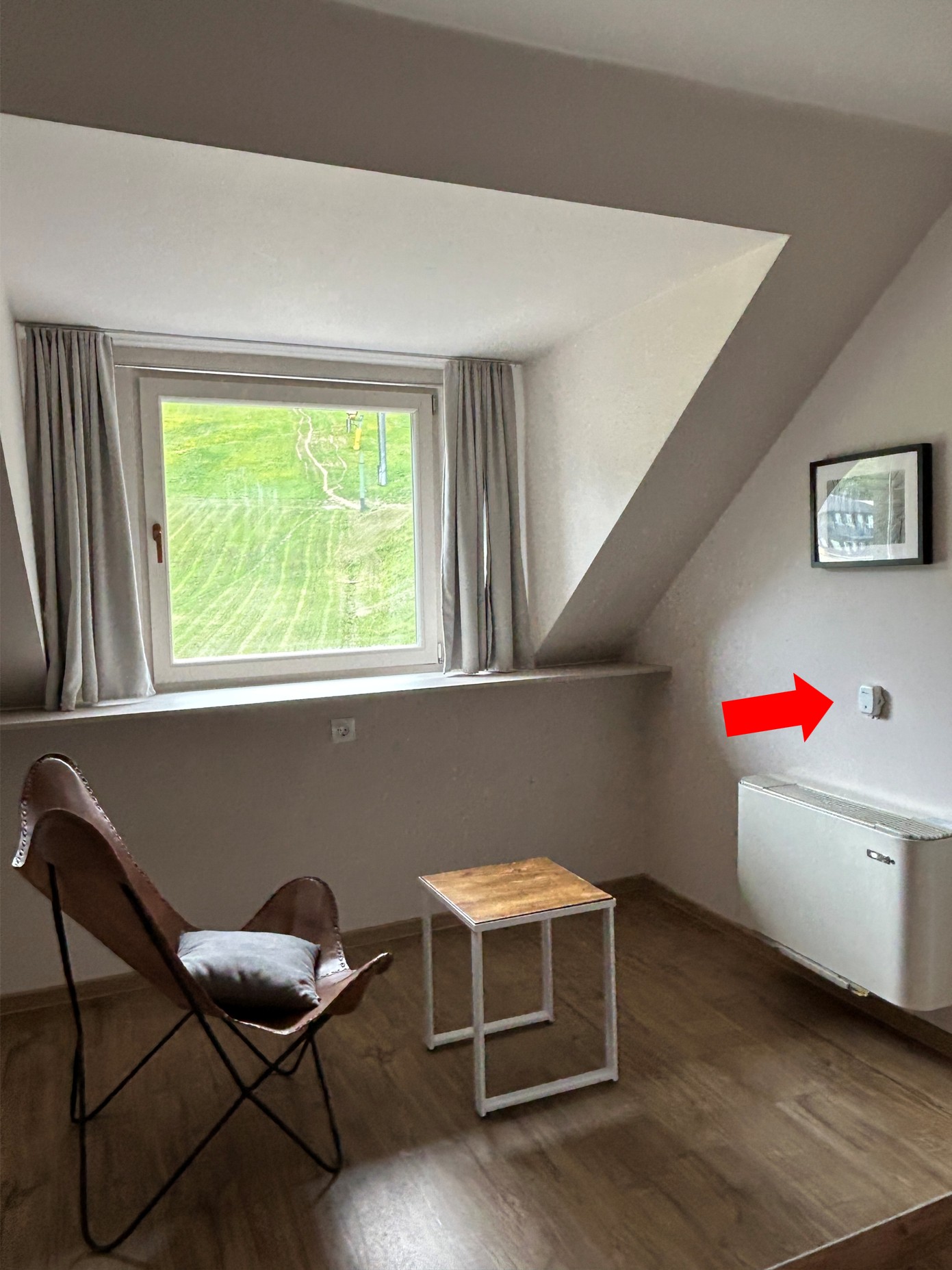}
    \caption{}
    \label{fig:install_guest_room}
  \end{subfigure}\hfill
  \begin{subfigure}[t]{0.24\textwidth}
    \includegraphics[width=\linewidth]{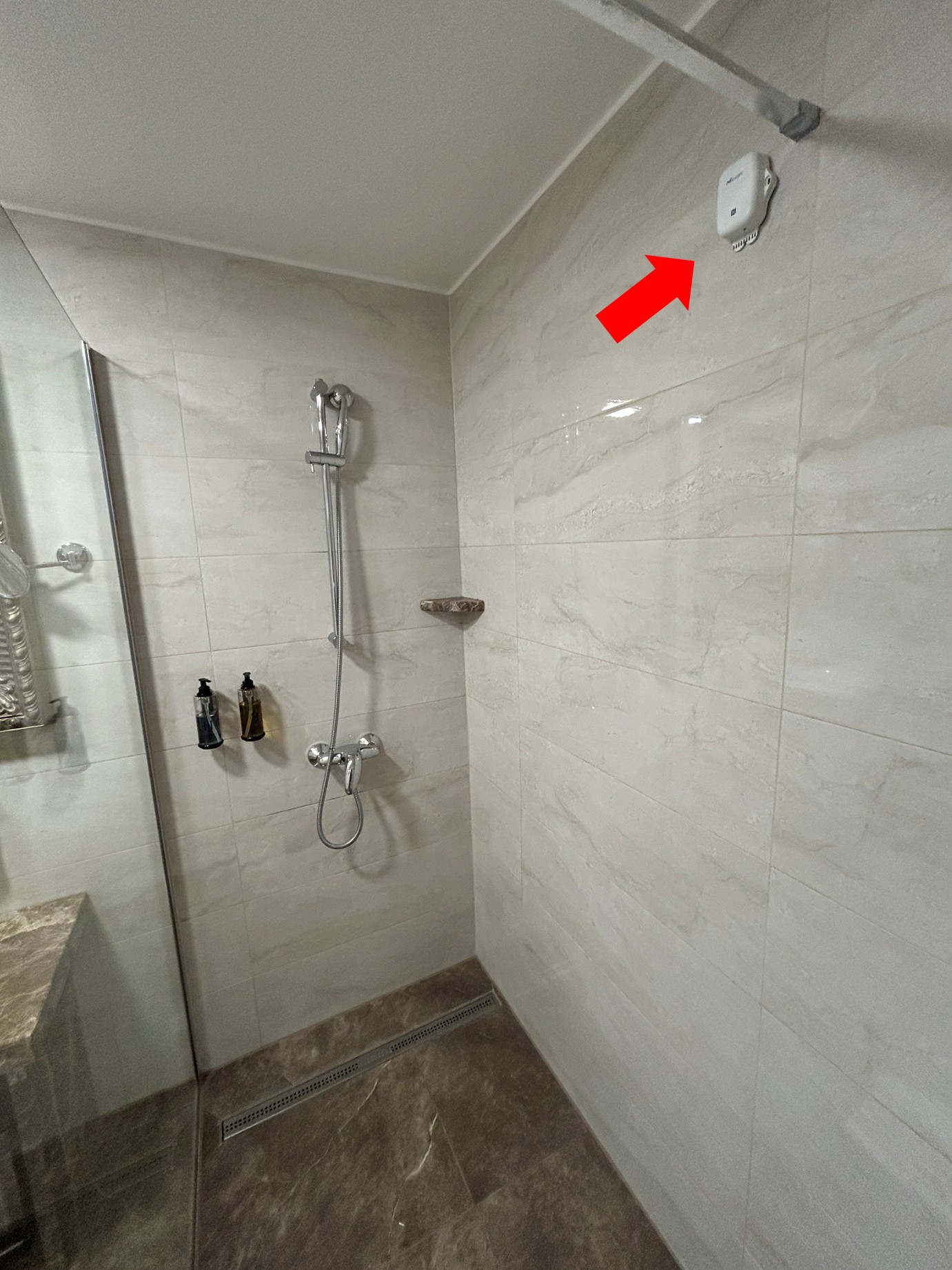}
    \caption{}
    \label{fig:install_bathroom}
  \end{subfigure}\hfill
  \begin{subfigure}[t]{0.24\textwidth}
    \includegraphics[width=\linewidth]{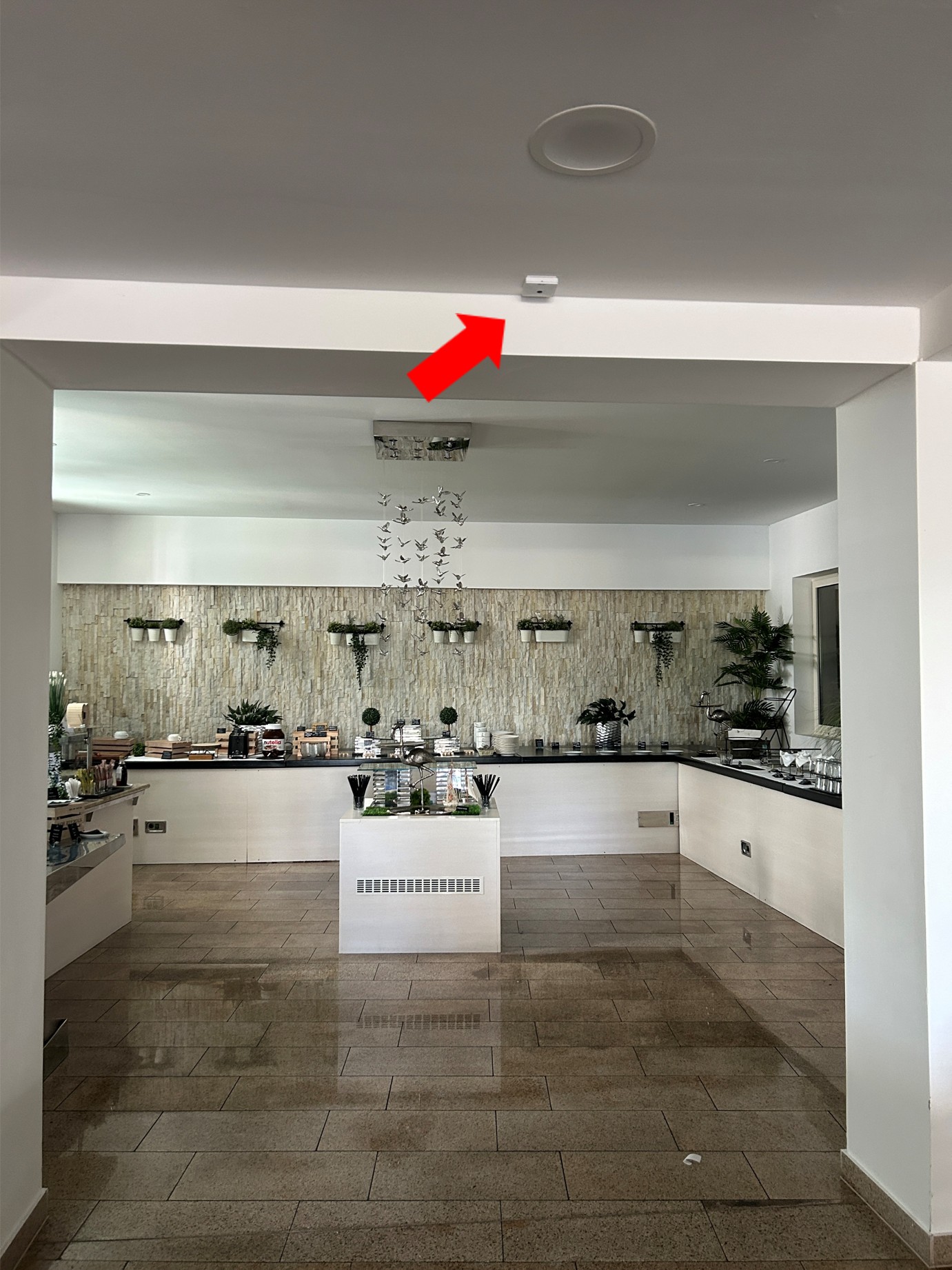}
    \caption{}
    \label{fig:install_dining}
  \end{subfigure}\hfill
  \begin{subfigure}[t]{0.24\textwidth}
    \includegraphics[width=\linewidth]{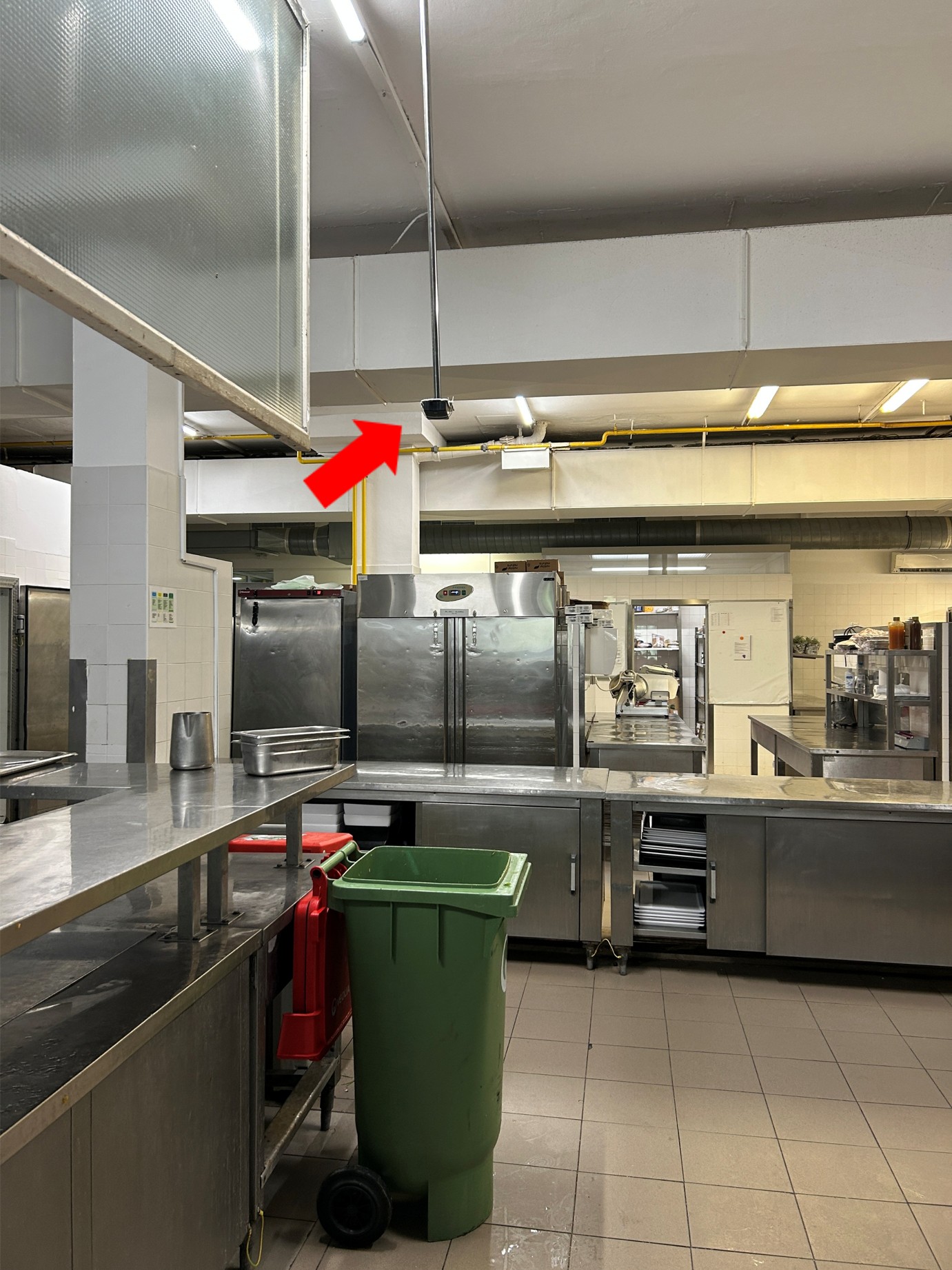}
    \caption{}
    \label{fig:install_kitchen}
  \end{subfigure}
  \caption{Example sensor installations: (a) EM300-TH temperature and humidity sensor in a guest room, e.g., for air conditioner runtime estimation, (b) EM300-TH mounted in shower cubicle for shower event detection, (c) VS121 people-counting sensor ceiling-mounted in a hotel dining hall, and (d) EM400-TLD fill level detection sensor in a hotel kitchen for monitoring food waste.}  \label{fig:installations}
\end{figure}

Fig.~\ref{fig:installations} shows representative installations across these space types.

\begin{table}[htbp]
\centering
\footnotesize
\setlength{\tabcolsep}{5pt}
\renewcommand{\arraystretch}{1.2}
\caption{Per-site deployment summary, showing country, site setting and location, sensor types deployed, and total sensor count. Commercial site names are withheld to protect deployment partners.}
\label{tab:deployment}
\begin{tabularx}{\textwidth}{@{}l >{\hsize=.85\hsize\raggedright\arraybackslash}X >{\hsize=1.15\hsize\raggedright\arraybackslash}X c@{}}
\toprule
\textbf{Country} & \textbf{Site / location} & \textbf{Sensor types} & \textbf{Sensors} \\
\midrule
\multirow{10}{*}{Australia}
 & Student accommodation, UQ, Brisbane & WS101, R809A, EM300-TH, W1, custom scale, VS121 & 235 \\ 
 & Airport motel, Brisbane        & EM300-TH, Smart CT10x, W1                        & 101 \\ 
 & Suburban hotel, Brisbane       & EM300-TH, WS301, R809A                           & 57  \\ 
 & Beachfront resort, Gold Coast  & EM300-TH, R809A, WS301, EM400-TLD, custom scale, VS121 & 38 \\ 
 & City hotel, Cairns             & EM300-TH, R809A, WS301, VS121, custom scale      & 23  \\ 
 & City hotel, Melbourne          & EM300-TH                                         & 23  \\ 
 & City hotel, Hobart             & R809A, EM300-TH, WS301                           & 20  \\ 
 & City hotel, Launceston         & WS301, R809A, EM300-TH                           & 17  \\ 
 & Home test site, Brisbane\textsuperscript{a} & AM319, WS523, WS301, ERS-Sound      & 14  \\ 
 & Lab test site, UQ, Brisbane\textsuperscript{a} & EM300-TH, WS156, WS101, AM319, WS523, ERS-Sound, WS301, EM310-UDL, Smart CT10x, W1, UC100, custom scale, EM400-TLD, R718N1, R211, UC300, WS201, EM400-MUD, EM400-UDL & 119 \\ 
\midrule
\multirow{9}{*}{Slovenia}
 & Countryside estate hotel, Kranj & EM300-TH, VS121, UC300           & 35 \\ 
 & Ski resort, Maribor            & EM300-TH, UC300                   & 34 \\ 
 & Beachfront resort, Portorož    & EM300-TH, custom scale, EM400-TLD & 34 \\ 
 & Seaside hotel, Portorož        & EM300-TH, UC300, VS121            & 33 \\ 
 & Student accommodation, Portorož & EM300-TH                         & 31 \\ 
 & Old-town boutique hotel, Piran & EM300-TH                          & 21 \\ 
 & Seaside hotel A, Piran         & EM400-TLD                         & 1  \\ 
 & Seaside hotel B, Piran         & EM400-TLD                         & 1  \\ 
 & Test and staging site\textsuperscript{a} & UC300, custom scale, VS121, EM300-TH, WS301, WS101 & 13 \\ 
\midrule
\textbf{Total} & & & \textbf{850} \\
\bottomrule
\multicolumn{4}{@{}p{\textwidth}@{}}{\footnotesize\textsuperscript{a}The 21 deployment sites appear as 19 rows. Five internal test, staging, and commissioning locations are consolidated into three rows, two in Australia and one in Slovenia. Sensor counts and sensor types in these rows are aggregated across the merged locations.} \\
\end{tabularx}
\end{table}

We grew the deployment incrementally from an initial pilot to its current scale, adding sites as partner agreements were established and refining installation procedures, decoder coverage, and dashboard configurations through each engagement. Each site has been monitored continuously since commissioning, with battery replacements, sensor maintenance, and any decoder updates coordinated with hotel housekeeping schedules to minimise disruption to guest operations.

\subsection{Collected sensor data}
\label{sec:data}

\begin{figure}[H]
  \centering
  \includegraphics[width=\columnwidth]{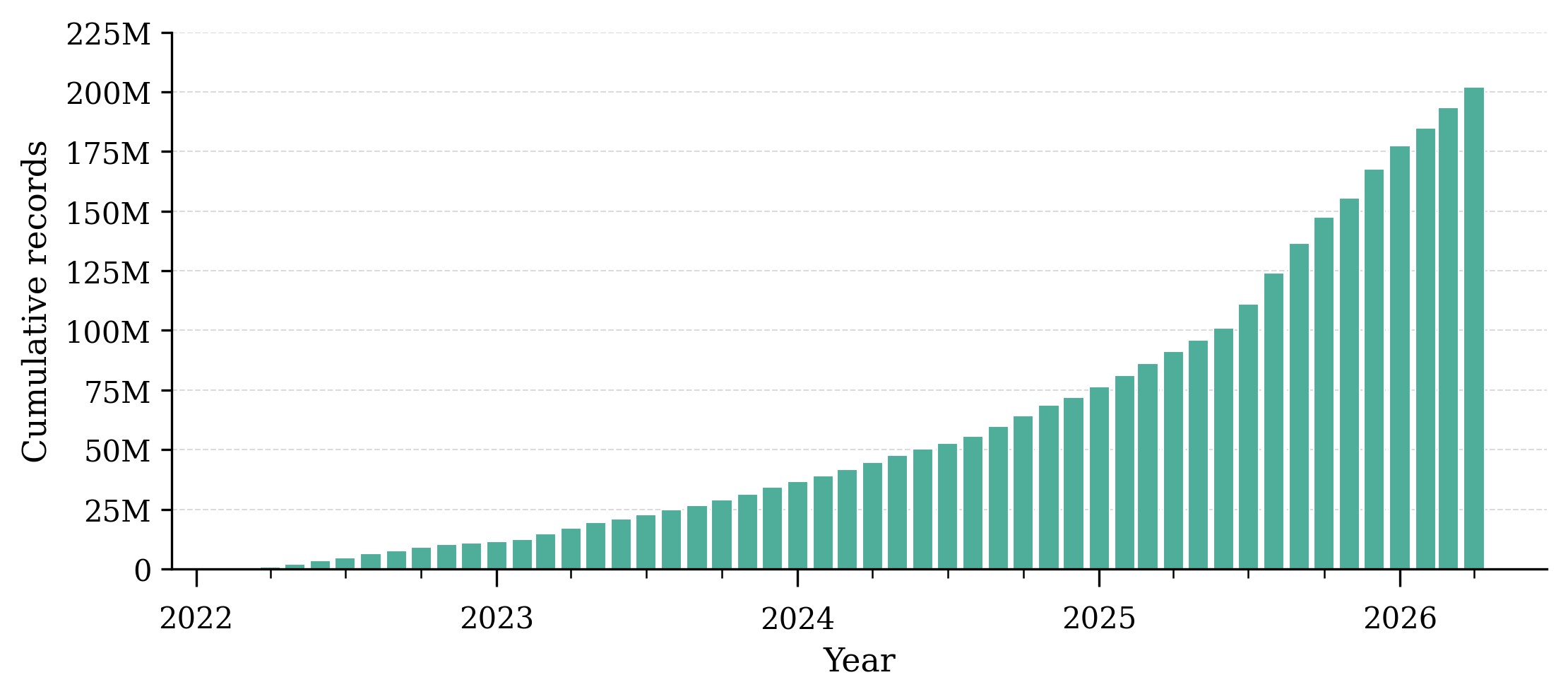}
  \caption{Cumulative record count over the deployment period, reaching approximately 202 million by April 2026. Growth steepens sharply from mid-2025 onwards, reflecting the period during which the deployment reached its current scale.}
  \label{fig:cumulative_volume}
\end{figure}

The system has run continuously since February 2022 and had accumulated approximately 202 million records as of April 2026, indexed in Elasticsearch across all deployed sensor types. Fig.~\ref{fig:cumulative_volume} shows the cumulative record count over the deployment period. Growth was driven by two factors: incremental site additions as partner agreements were established, and the higher reporting frequency of the EM300-TH temperature and humidity sensors (one-minute intervals) compared to other sensor types in the deployment. The EM300-TH alone accounts for roughly 90\% of the total record count despite making up under half of deployed sensors, a direct consequence of the one-minute reporting choice discussed in Section~\ref{sec:sensors}.
Each record carries one or more decoded modalities, ranging from a single weight measurement on the custom platform scales to nine simultaneous environmental readings on the Milesight AM319. Across the full dataset, the 202 million records correspond to approximately 453 million individual sensor readings. Temperature and humidity dominate this count, contributing 189 million and 185 million readings respectively, together approximately 82\%. The 4 million difference between them reflects the sensors that report temperature without humidity, including the EM310-UDL, EM400-X, WS201, and WS301. The remaining 80 million readings span 26 further modalities including indoor air quality (carbon dioxide, total volatile organic compounds, and particulate matter), acoustic levels, electrical consumption, water flow and fill level, weight, occupancy and motion, and discrete events.

To our knowledge, this is one of the largest LoRaWAN datasets generated for hotel sustainability research, both in record count and in deployment duration. For comparison, Rosa et al.~\cite{rosa_development_2026} report 5.4 million records over one year across three dairy farms; our dataset is approximately 37 times larger over a four-year period and spans a wider mix of sensor types and operational contexts.
The cumulative count (Fig.~\ref{fig:cumulative_volume}) reaches roughly 75 million records by the start of 2025 and a further 125 million in the 16 months that follow, reflecting both the onboarding of larger sites and the matured reporting cadence of the EM300-TH fleet.

\subsection{Known data gaps}
\label{sec:known_gaps}
Data gaps are identifiable through frame counter discontinuities and timestamp gaps, allowing downstream analyses to either exclude affected periods or apply interpolation as appropriate. The largest gap in the dataset to date occurred from 12 January to 15 March 2026, when a gateway was inadvertently unplugged at one site, halting uplinks from the 235 sensors served by that gateway until the connection was restored. Smaller intermittent gaps occur during scheduled maintenance, hotel-side network reconfiguration, and severe weather events that disrupt backhaul connectivity. These patterns are consistent with reports from other long-running IoT deployments~\cite{rosa_development_2026}, where backhaul and power interruptions are the dominant sources of data loss rather than radio-layer failures.

A small number of records carry malformed timestamps from sensor or gateway clock faults (e.g., uplinks stamped as 2105 rather than 2025). These are filtered from analyses by restricting the time range to the deployment window. Their existence does not affect record counts at the resolution reported here, but downstream studies that depend on exact event timing should apply the same filter at query time.

Sensor-level losses also occur when guests remove sensors from guest rooms, occasionally discarding them or taking them away. These events surface in the data as an abrupt, permanent stop in a single sensor's uplinks while neighbouring sensors continue reporting, which distinguishes them from gateway or backhaul outages that silence many sensors at once. We identify affected sensors through this signature and replace them during the next maintenance visit.

\section{LoRaWAN message delivery reliability}
\label{sec:reliability}

Before evaluating message delivery, the principal LoRaWAN communication parameters used across the deployment are summarised in Table~\ref{tab:lorawan_configuration}. The Milesight sensors allow several parameters to be configured, including activation mode, adaptive data rate (ADR), uplink confirmation, spreading factor, transmission power, channel selection within the supported regional band, and reporting interval. These parameters influence communication reliability, network airtime, gateway capacity, and sensor battery consumption.

In this deployment, ADR was enabled, allowing the network server to adjust the spreading factor and transmission power according to link conditions. The transmission bandwidth was fixed at 125~kHz and was not adjusted by ADR. Confirmed uplinks (\texttt{ConfirmedDataUp}) were also enabled. Therefore, the network server returned an acknowledgement for each successfully received uplink; when an acknowledgement was not received, the Milesight sensor retransmitted the message once. This configuration was intended to improve message delivery reliability, although acknowledgements and retransmissions increase downlink traffic, channel occupancy, and battery consumption.

The sensors deployed in Australia operated in the AU915--928~MHz regional frequency band. The supported regional band is determined by the sensor hardware and cannot be changed through software configuration. Consequently, deployment in another regulatory region requires the corresponding hardware variant; for example, European sensors operate in the EU863--870~MHz band, commonly referred to as EU868. Individual channels may be enabled or disabled within the sensor's supported regional band, but the sensor cannot be reconfigured to operate in a different regional frequency band.

\begin{table}[htbp]
\centering
\caption{Principal LoRaWAN parameters used across the Australian and Slovenian deployments.}
\label{tab:lorawan_configuration}
\begin{tabularx}{\linewidth}{>{\raggedright\arraybackslash}p{0.28\linewidth} >{\raggedright\arraybackslash}p{0.28\linewidth} >{\raggedright\arraybackslash}X}
\toprule
\textbf{Parameter} & \textbf{Configuration} & \textbf{Explanation} \\
\midrule
Activation mode & OTAA & The network server derives fresh session keys at each join. \\
Device class & A and C & Class A: battery-powered \newline Class C: mains-powered \\
Regional band & AU915--928~MHz \newline EU863--870~MHz & Australia \newline Slovenia \\
Bandwidth & 125~kHz & Fixed for all uplinks. ADR does not adjust it. \\
Spreading factor & Set by ADR & ADR, enabled on all sensors, dynamically selects optimal SF. \\
Transmission power & Set by ADR & Network server adapts power based on link quality. \\
Uplink message type & Confirmed Data Up & Unacknowledged uplink messages are retransmitted. \\
Payload size & 7--11~bytes & Depending on sensor telemetry data. \\
\bottomrule
\end{tabularx}
\end{table}

The reliability of the deployed LoRaWAN links was evaluated using packet delivery ratio (PDR), signal-to-noise ratio (SNR), and spreading factor (SF) usage. These indicators provide complementary information about network performance: PDR quantifies the proportion of expected packets that were successfully received, SNR describes the quality of the received radio signal relative to the background noise, and SF reflects the adaptive LoRaWAN transmission configuration used to maintain communication under different link conditions.

\begin{figure}[H]
    \centering
    \includegraphics[width=0.75\linewidth]{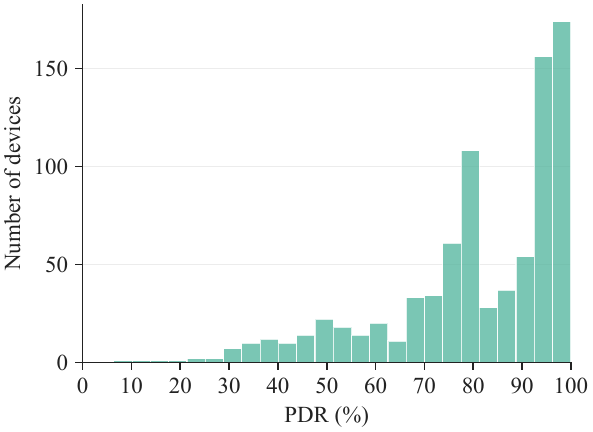}
    \caption{Distribution of packet delivery ratio (PDR) across deployed LoRaWAN sensors.}
    \label{fig:pdr_histogram}
\end{figure}

Across the deployment, 831 sensors were retained after data cleaning, contributing 184,775,002 received packets. The mean PDR was 80.9\%, with a median of 85.4\% and a standard deviation of 18.3 percentage points, indicating generally high delivery reliability but with substantial device-level variation. The PDR distribution in Fig.~\ref{fig:pdr_histogram} shows that most sensors achieved relatively high packet delivery, although a smaller subset of sensors experienced noticeably lower reliability. This spread is expected in real-world IoT deployments because link quality can vary with sensor placement, building layout, gateway distance, obstruction, antenna orientation, and local interference~\cite{teymuri_lp-mab_2023}.

\begin{figure}[!t]
    \centering
    \includegraphics[width=0.75\linewidth]{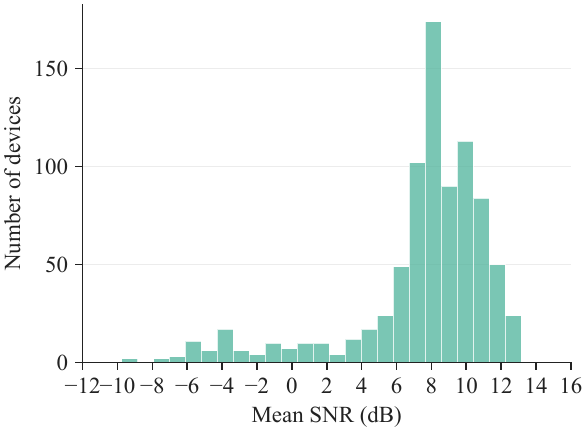}
    \caption{Distribution of mean signal-to-noise ratio (SNR) across deployed LoRaWAN sensors.}
    \label{fig:snr_histogram}
\end{figure}

The SNR distribution is shown in Fig.~\ref{fig:snr_histogram}. The mean SNR was 7.5~dB and the median was 8.2~dB, with a standard deviation of 4.0~dB. Negative SNR values are not necessarily invalid in LoRaWAN systems, as LoRa modulation can still decode packets below the noise floor; however, such values typically indicate a more challenging communication environment and can be associated with reduced reliability or greater dependence on robust transmission settings.

\begin{figure}[!t]
    \centering
    \includegraphics[width=0.95\linewidth]{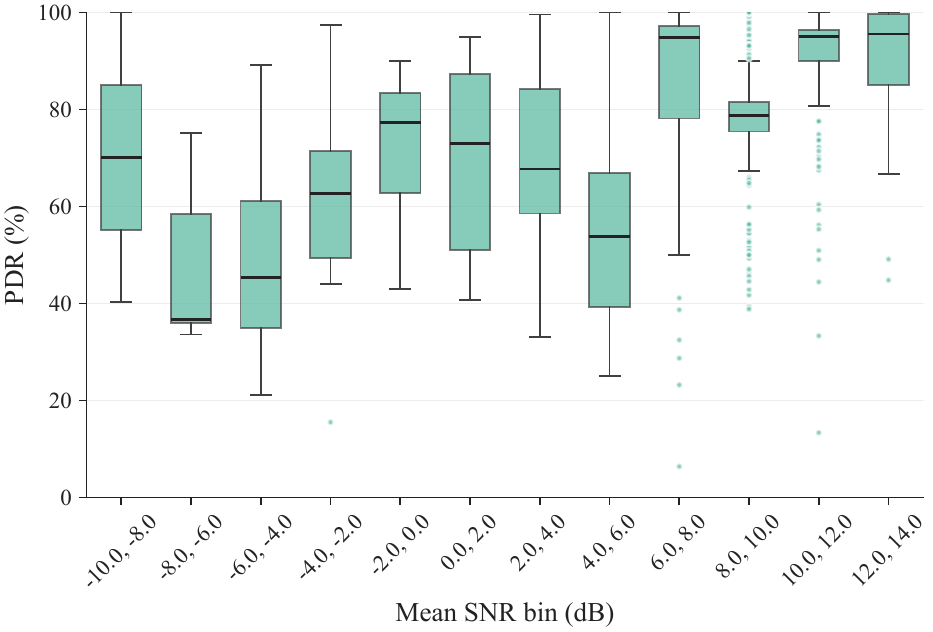}
    \caption{Packet delivery ratio (PDR) grouped by mean SNR bins. Each box represents the sensor-level PDR distribution within a 2~dB SNR interval.}
    \label{fig:pdr_snr_boxplot}
\end{figure}

To further examine the relationship between link quality and delivery reliability, PDR was grouped in 2~dB SNR intervals, as shown in Fig.~\ref{fig:pdr_snr_boxplot}. The boxplot shows that sensors with very low SNR generally had lower and more variable PDR. For example, sensors in the $[-8,-6)$~dB and $[-6,-4)$~dB SNR bins had mean PDR values of 47.9\% and 49.2\%, respectively. In contrast, sensors with stronger SNR generally achieved higher delivery reliability. Devices in the $[10,12)$~dB and $[12,14)$~dB bins had mean PDR values of 90.8\% and 90.1\%, respectively. This trend confirms that improved SNR is broadly associated with stronger message delivery reliability.

However, the relationship is not perfectly monotonic. Some mid-to-high SNR bins still contain low-PDR outliers, suggesting that SNR alone does not fully explain packet loss. This is important for an IoT deployment because message delivery can also be affected by gateway congestion, temporary outages, duty-cycle limitations, sensor firmware behaviour, battery state, packet collisions, local interference, and site-specific installation conditions. Therefore, while SNR is a useful indicator of radio-link quality, PDR remains the more direct measure of end-to-end sensing-data availability.

Table~\ref{tab:lorawan_delivery_summary} summarises the key deployment, PDR, and SNR statistics. These results show that the dataset provides a large-scale empirical basis for assessing LoRaWAN reliability across many sensors and rooms. The high median PDR indicates that most sensors delivered a substantial proportion of their expected packets, while the standard deviation highlights the need to account for sensor-level heterogeneity when using the data for downstream modelling or monitoring.

\begin{table}[!t]
\centering
\caption{Summary statistics for LoRaWAN delivery reliability and link quality across 831 sensors and 184,775,002 received packets.}
\label{tab:lorawan_delivery_summary}
\begin{tabular}{lrrrr}
\toprule
\textbf{Metric} & \textbf{Mean} & \textbf{Median} & \textbf{SD} & \textbf{Maximum} \\
\midrule
Active days & 584.7 & 533 & & 1,452 \\
Packet delivery ratio (\%) & 80.9 & 85.4 & 18.3 & \\
Signal-to-noise ratio (dB) & 7.5 & 8.2 & 4.0 & \\
\bottomrule
\end{tabular}
\end{table}

The distribution of received packets across spreading factors is reported in Table~\ref{tab:sf_distribution}. Network traffic was strongly dominated by SF7, which accounted for 178,022,485 packets, equivalent to 96.35\% of all received packets. SF8 contributed a further 1.59\%, while each of SF9--SF12 accounted for less than 1\% of the total. This indicates that most transmissions used the lowest spreading factor, which provides a higher data rate and shorter time-on-air. Only a small proportion of traffic used higher spreading factors, suggesting that more robust transmission settings were required relatively infrequently.

The mean PDR associated with SF7 was 81.22\%, whereas SF8 and SF9 had lower mean PDR values of 64.66\% and 54.87\%, respectively. This pattern suggests that higher spreading factors were generally used under more challenging radio conditions. SF10 had a particularly low mean PDR of 15.54\%. One possible explanation is that much of the SF10 traffic occurred during temporary ADR adjustment or connectivity-recovery periods, rather than representing a stable long-term transmission setting. Under ADR, ordinary uplink packets are used by the network server to evaluate link quality and determine suitable transmission parameters. If connectivity deteriorates or downlinks are not received, the sensor can progressively move to a lower data rate and higher spreading factor to recover the link. Therefore, packets observed at SF10 may represent a transitional stage while the sensor was moving towards a more robust setting. 

\begin{table}[!t]
\centering
\caption{Spreading factor distribution and associated LoRaWAN reliability statistics.}
\label{tab:sf_distribution}
\small
\setlength{\tabcolsep}{4pt}
\begin{tabular}{lrrrr}
\hline
\textbf{SF} & \textbf{Packets} & \textbf{Packet share (\%)} &
\textbf{Mean PDR (\%)} & \textbf{Mean SNR (dB)} \\
\hline
SF7  & 178,022,485 & 96.35 & 81.22 & 7.75 \\
SF8  & 2,945,411   & 1.59  & 64.66 & -3.17 \\
SF9  & 1,553,630   & $<1$  & 54.87 & -5.24 \\
SF10 & 1,117,083   & $<1$  & 15.54 & -3.85 \\
SF11 & 447,747     & $<1$  & 82.40 & -0.90 \\
SF12 & 688,646     & $<1$  & 84.54 & -0.12 \\
\hline
\textbf{Total} & \textbf{184,775,002} & \textbf{100.00} & & \\
\hline
\end{tabular}
\end{table}

Overall, the results indicate that the LoRaWAN deployment achieved generally reliable message delivery, with most received traffic transmitted efficiently using SF7. At the same time, the variation in PDR across spreading factors highlights the heterogeneous communication conditions present in a real-world building deployment. Low-PDR observations and the limited traffic at higher spreading factors emphasise the importance of validating communication reliability before using IoT sensor streams for building monitoring, inference, or operational decision-making. These reliability indicators provide important context for the subsequent analyses because downstream model performance depends not only on the measured variables, but also on the consistency with which sensor messages are delivered and the long-term availability of battery-powered sensors.

\section{Sensor battery life}

In addition to message-delivery reliability, we examined battery behaviour to assess the operational lifetime of the deployed sensors. We based this analysis on 13 Milesight EM300-TH temperature and humidity sensors at the Beachfront resort, Gold Coast site. We selected this set because the sensors were deployed together, carried the same battery type, ran the same firmware, and shared a one-minute reporting interval, so any differences in battery behaviour reflect the sensors themselves rather than differences in configuration or deployment timing.

\begin{figure}[H]
    \centering
    \includegraphics[width=0.90\linewidth]{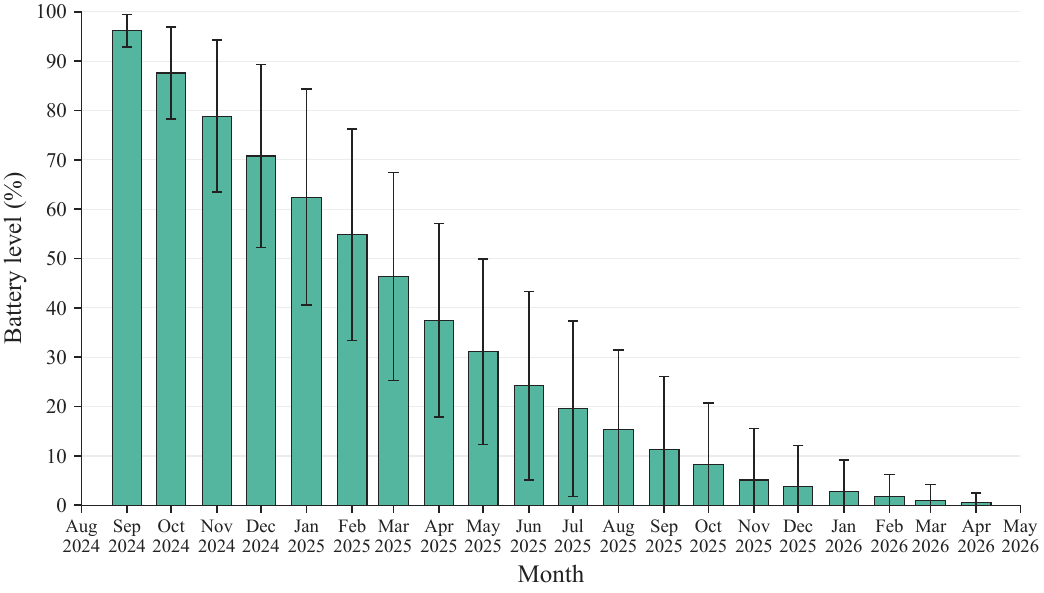}
    \caption{Monthly mean battery level across 13 Milesight EM300-TH temperature and humidity sensors deployed at the Beachfront resort, Gold Coast, each reporting at one-minute intervals. Error bars indicate one standard deviation across sensors within each month.}
    \label{fig:battery_depletion}
\end{figure}

Fig.~\ref{fig:battery_depletion} presents the monthly mean battery level across the 13 sensors. Error bars represent one standard deviation across sensors within each month and therefore indicate variation in battery depletion rates between sensors. The results show a clear and sustained decline in battery level over the deployment period. The monthly mean was approximately 95\% at the beginning of monitoring in September 2024 and decreased to approximately 70\% by December 2024. It fell below 50\% by March 2025, below 20\% by July 2025, and approached zero during early 2026. This trend indicates that most EM300-TH sensors reached the end of their effective battery life after approximately 18--20 months of operation at a one-minute reporting interval, the most demanding cadence in the deployment; sensors reporting at 10--15 minute intervals last correspondingly longer.

The relatively large error bars during the middle of the deployment indicate noticeable variation in battery depletion across sensors, despite all sensors being the same model and configured to transmit at one-minute intervals. This variation is therefore more likely to reflect differences in radio-link conditions, retransmission activity, spreading factor, transmission power, local environmental exposure, and unit-to-unit hardware or battery variability. Some sensors consequently reached low battery levels earlier than others, while others retained higher battery levels for longer. Battery condition should therefore be considered alongside PDR and SNR when evaluating the long-term reliability and maintenance requirements of LoRaWAN sensing deployments.

\section{Use cases}
\label{sec:use_cases}

The system has supported seven field studies testing the effectiveness of behaviour change interventions across food waste, energy consumption, and water consumption (Table~\ref{tab:use_cases}). The studies share the same sensing infrastructure, but each uses a different combination of sensors, reporting intervals, and inference methods. Together they show the breadth of the platform and the design patterns that recur when sensing for sustainability in operational hotels.

\begin{table}[H]
\centering
\caption{Field studies and interventions enabled by the platform, with sensors used and source publications.}
\label{tab:use_cases}
\begin{tabular}{p{3.8cm} p{4.2cm} p{4.5cm}}
\toprule
\textbf{Use case} & \textbf{Publication} & \textbf{Sensors used} \\
\midrule
\multicolumn{3}{l}{\textit{Waste monitoring}} \\
\midrule
Automated plate waste measurement & Dolnicar et al.~\cite{dolnicar_automatically_2023} & Custom scale \\
Table sign intervention & Zinn et al.~\cite{zinn_not_2026} & Custom scale \\
\addlinespace
\multicolumn{3}{l}{\textit{Energy consumption}} \\
\midrule
Air conditioning use & Greene et al.~\cite{greene_leveraging_2025} & EM300-TH \\
Heater use & Greene et al.~\cite{greene_crikey_2025} & EM300-TH \\
Minibar energy usage & Dolnicar et al.~\cite{dolnicar_does_2024} & WS523, R809A, WS301 \\
Air-conditioner runtime estimation (ML) & Serati et al.~\cite{serati_ac_runtime} & EM300-TH \\
\addlinespace
\multicolumn{3}{l}{\textit{Water consumption}} \\
\midrule
Shower duration estimation (ML) & Sablica et al.~\cite{sablica_ecoshower_2025} & EM300-TH, ERS-Sound, WS101 \\
\bottomrule
\end{tabular}
\end{table}

\subsection{Waste monitoring}

Dolnicar et al.~\cite{dolnicar_automatically_2023} used the platform to measure plate waste at a hotel buffet. They placed a commercial steel-framed load cell scale (150 kg capacity, 10 g precision) beneath the kitchen waste bin, connected through the RS232-to-LoRaWAN bridge described in Section~\ref{sec:sensors}. Combining the maximum daily plate weight with the daily guest count supplied by hotel management gives plate waste per guest, the primary metric for intervention studies. The paper illustrates the approach with three weeks of measurements, with typical values of 200--300 g per person per day.

A subsequent study used the same sensing setup at another property for a controlled field experiment. Zinn et al.~\cite{zinn_not_2026} ran a five-condition study at an Australian hotel across 176 usable days, combining a plate-waste fact with ascription-of-responsibility, habit-transfer, or effort-reduction framings, plus a control. A tablet at the buffet entrance captured parallel satisfaction ratings. No condition significantly reduced plate waste relative to the control condition ($F(3, 140) = 0.41$, $p = 0.744$; means ranged from 73.7 to 80.0 g per guest), and two conditions (the plate-waste fact alone and the fact-plus-effort framing) produced small but statistically significant decreases in guest satisfaction. Automated continuous measurement combined with parallel satisfaction monitoring made it possible to evaluate not just whether the interventions reduced waste, but whether they reduced guest experience without delivering environmental benefits.

\subsection{Energy consumption}

Greene et al.~\cite{greene_leveraging_2025} used the platform to study guest air-conditioning behaviour in summer, deploying EM300-TH temperature and humidity sensors in 24 rooms of a small midrange Brisbane hotel across seven experimental conditions (control, basic instructions, social norms, and four anthropomorphism interventions personifying the air conditioner with different emotional expressions). Daily average room temperature served as a proxy for sustainable cooling behaviour: in summer, a higher room temperature means less overcooling. After controlling for outdoor temperature and pre-intervention room baseline (1,361 observations across 75 days), four conditions significantly increased room temperature relative to the control condition: basic instructions ($+0.50$\textdegree C, $d = 0.45$), social norms ($+0.51$\textdegree C, $d = 0.47$), anthropomorphism-exhausted ($+0.31$\textdegree C, $d = 0.28$), and anthropomorphism-angry ($+0.28$\textdegree C, $d = 0.25$). The positive-emotion anthropomorphism conditions (calm, enjoyable) had no significant effect.

A companion winter study by Greene et al.~\cite{greene_crikey_2025} reversed the seasonal direction at the same Brisbane site, using the same EM300-TH sensors in 20 rooms to test five message conditions (control, basic instructions, environmental beliefs, compassion, humour) over July--August 2024. With lower room temperature now indicating more sustainable behaviour, all four message conditions significantly reduced room temperature relative to control, with effect sizes ranging from $d = 0.46$ (environmental beliefs) to $d = 0.62$ (humour). Together, the two studies demonstrate that the platform's one-minute reporting interval and continuous coverage support per-room daily averaging in both heating and cooling regimes, with the same sensor model and analytic design generalising across seasons.

\begin{figure}[H]
  \centering
  \includegraphics[width=\columnwidth]{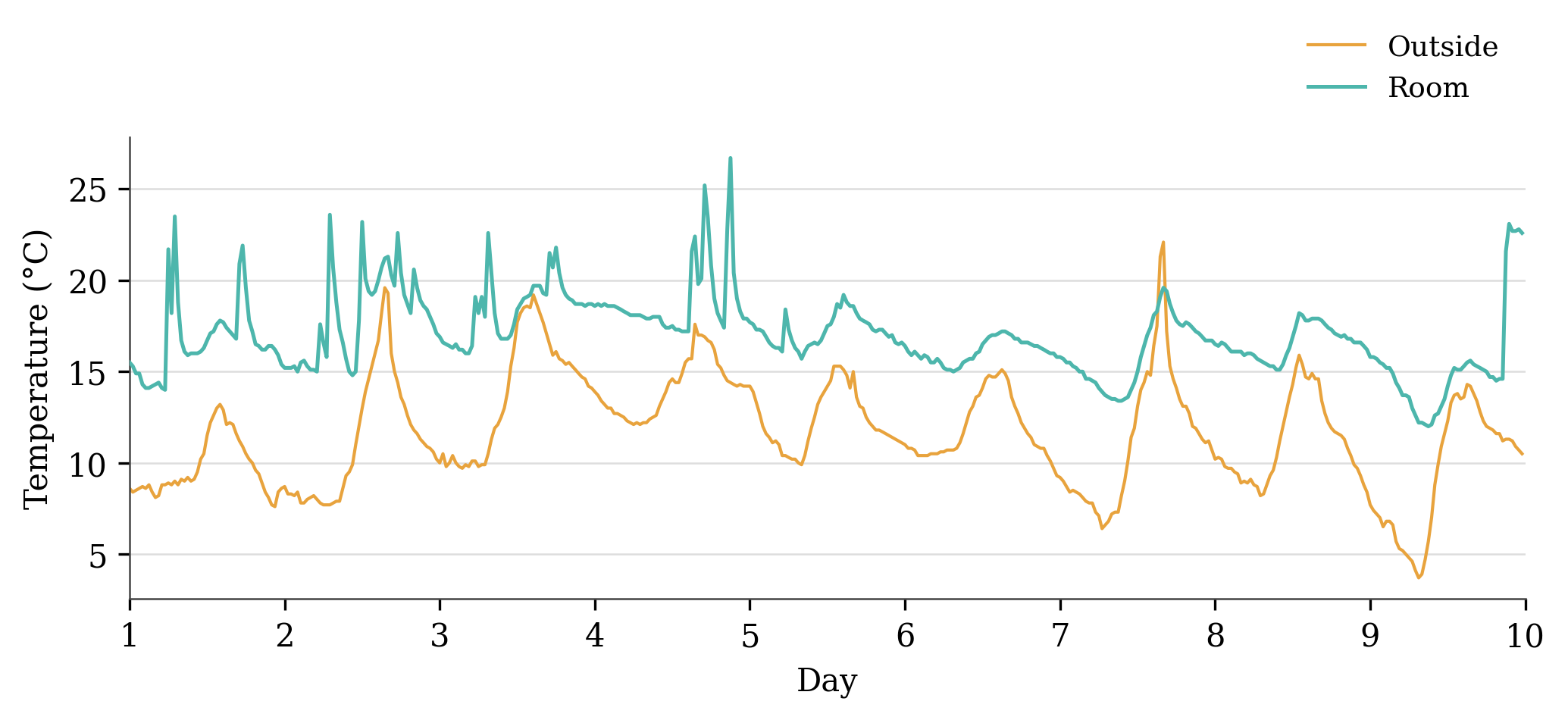}
  \caption{Hourly outdoor temperature (orange) and room temperature (teal) for one instrumented room at a cool-climate site over one week. The room is held well above outdoor temperature throughout, with short spikes to 23--27\textdegree C where the occupant raised the setpoint above a comfortable level.}
  \label{fig:room_temp_trace}
\end{figure}

Beyond these intervention studies, the platform's per-room temperature traces make individual heating and cooling behaviour directly visible. Fig.~\ref{fig:room_temp_trace} shows hourly outdoor temperature against room temperature for one instrumented room at a cool-climate site over one week. The outdoor temperature follows a clear diurnal cycle between roughly 4 and 20\textdegree C, while the room stays consistently warmer, mostly between 15 and 20\textdegree C, as the room is actively heated. Short upward spikes to 23--27\textdegree C recur through the week, marking periods when the occupant raised the air conditioner temperature setpoint well above a comfortable level. Traces of this kind are the raw signal underlying the daily room-temperature averages used in the intervention studies above, and they show how individual setpoint behaviour, not just outdoor conditions, drives in-room energy use.

Dolnicar et al.~\cite{dolnicar_does_2024} used the platform to measure minifridge energy consumption and guest use across 19 minifridges in four hotels (two in Tasmania, two in Southeast Queensland) over four months. Each minifridge was instrumented with a Netvox R809A smart power plug measuring cumulative energy consumption ($\pm$3\% accuracy) and a Milesight WS301 magnetic contact switch on the fridge door recording open/close events at one-second resolution; both sensors transmitted via LoRaWAN to the cloud infrastructure described in Section~\ref{sec:network_server}. The average daily energy consumption (539.83 Wh) was statistically independent of door-opening frequency ($r = 0.03$, $p = 0.245$), meaning the operational cost and carbon emissions of running each minifridge are incurred regardless of whether guests use it. A parallel survey of 400 respondents found that only 60\% of guests who had a minifridge in their room reported using it, with most expressing no strong negative reaction to the hypothetical removal of the minifridge or its provision only on request. The paper's contribution is methodological as much as substantive: it demonstrates how the platform supports studies combining continuous direct-energy measurement with parallel use-detection, enabling claims about resource use that single-sensor approaches could not support.

\begin{figure}[H]
  \centering
  \includegraphics[width=\columnwidth]{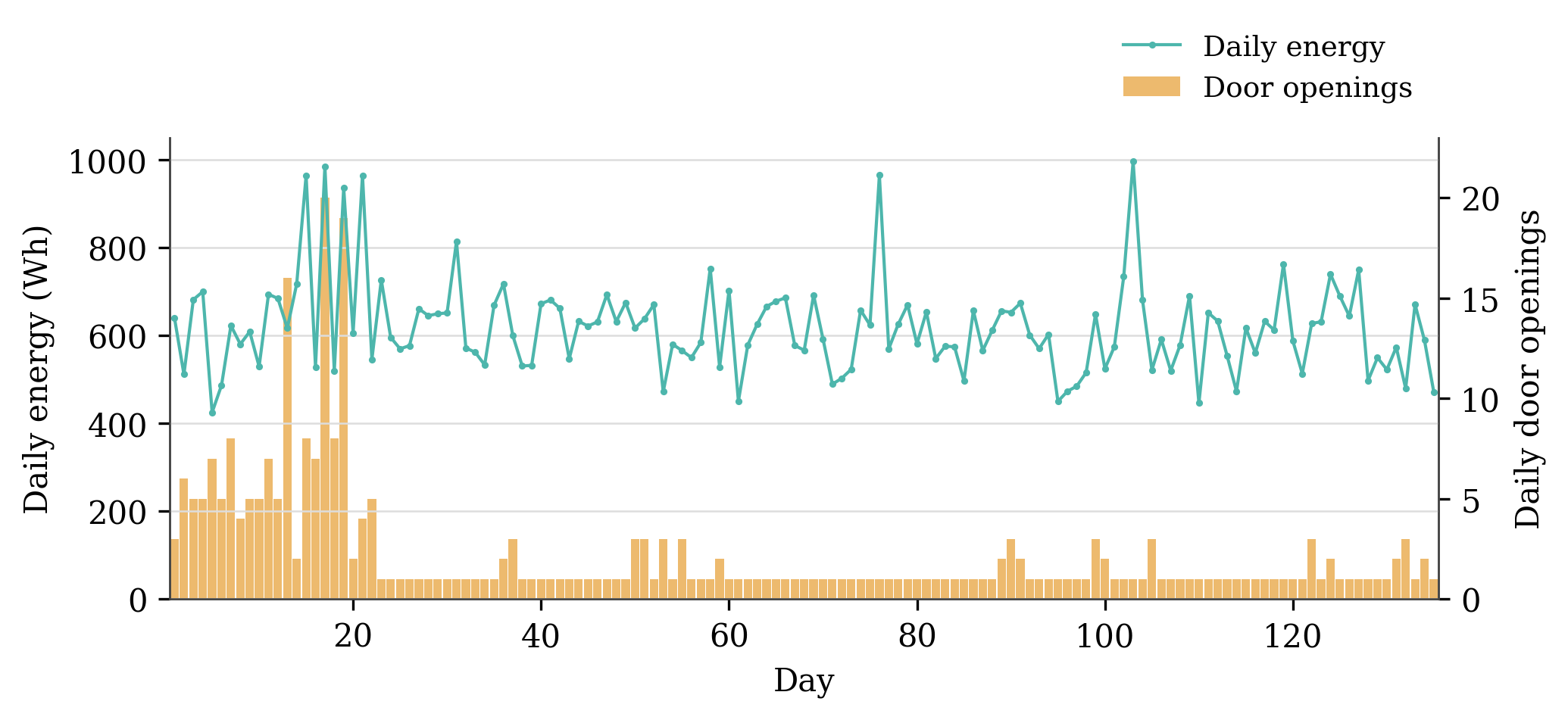}
  \caption{Daily minifridge energy consumption (teal line, left axis) against daily door-opening count (orange bars, right axis) for one instrumented room over a four-month period. On most days the minifridge is opened once, consistent with a single housekeeping check; higher counts indicate active guest use. Energy consumption remains within a narrow band irrespective of opening frequency, consistent with the deployment-wide finding that energy use is independent of guest use.}
  \label{fig:minibar_room617}
\end{figure}

Fig.~\ref{fig:minibar_room617} illustrates this pattern at room level using platform data from one instrumented room over a four-month period. The door-opening count separates two kinds of days: on most days the minifridge is opened once, consistent with a single housekeeping check, whereas the cluster of higher counts in the opening weeks reflects active guest use. Daily energy consumption holds within a narrow band (roughly 400--800~Wh) on both kinds of day, showing that the energy cost of running the minifridge persists whether or not guests use it.

Serati et al.~\cite{serati_deep_2026} used the platform to estimate room-level air-conditioner operation from indoor temperature and humidity measurements. The study drew on data from 17 hotel rooms, with independently established ON/OFF labels used to train and evaluate several deep-learning models, including Bidirectional Long Short-Term Memory (BiLSTM) and one-dimensional convolutional neural network (1D-CNN) architectures. The best-performing BiLSTM model achieved an F1-score of 97.5\% on an unseen hotel room, with a median daily runtime-estimation error of approximately 4 minutes. The BiLSTM model outperformed the 1D-CNN and the other architectures tested, demonstrating the platform's potential for non-intrusive air-conditioner monitoring where direct electricity measurements are unavailable.

\begin{figure}[H]
    \centering
    \includegraphics[width=0.9\columnwidth]{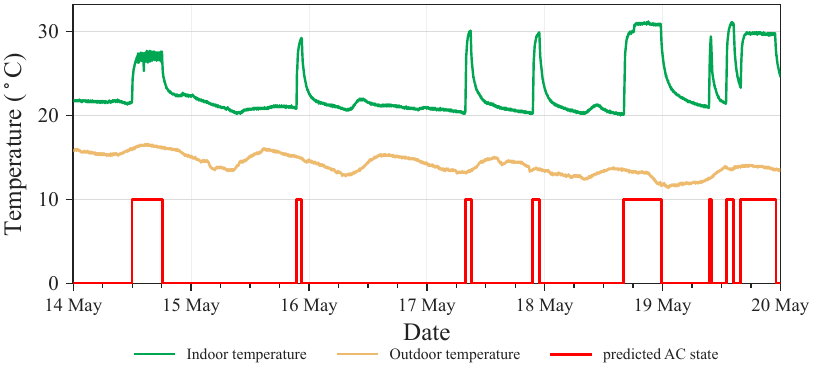}
    \caption{Indoor and outdoor temperatures together with the air-conditioner ON/OFF state predicted by the ML model.}
    \label{fig:ac_runtime_usecase}
\end{figure}

Fig.~\ref{fig:ac_runtime_usecase} shows one example week from the held-out hotel room. The green and orange lines give indoor and outdoor temperature, and the red shading marks the periods where the model predicts the air conditioner is on. The predicted operating periods generally coincide with drops or plateaus in indoor temperature, while outdoor temperature follows its own daily pattern. The week contains operating cycles of different lengths, showing that the model can identify both sustained and intermittent operation from changes in the room's ambient conditions. We scaled the binary model output vertically only to keep the predicted ON periods visible alongside the temperature traces.

\subsection{Water consumption}
Sablica et al.~\cite{sablica_ecoshower_2025} used the platform to develop and validate EcoShower, a non-intrusive shower-duration sensor based on humidity inference. They instrumented a bathroom in a residential property in Brisbane, Australia, pairing an Elsys ERS-Sound LoRaWAN sensor (average and peak sound), a Milesight EM300-TH sensor (temperature, humidity) with two Milesight WS101 smart buttons. The occupant pressed the buttons at the start and end of each shower to record ground truth. After cleaning, the analysis covered 30,645 twenty-second intervals across 227 days, split into 180 training days and 47 testing days. A BiLSTM classifier trained on all four sensor modalities achieved 97.1\% interval-level accuracy, predicting 704 shower-positive intervals against a ground truth of 746 over the 47-day test window (mean daily error 1.0~min, 2.2\%). A Gated Transformer Network performed comparably (96.5\% accuracy) at substantially higher computational cost. An ablation analysis showed that humidity data alone is sufficient: the humidity-only sub-model retained 96.0\% accuracy, which allows shower detection from a single Milesight EM300-TH (Section~\ref{sec:sensors}, Table~\ref{tab:sensors}) without adding sound, motion, or flow instrumentation to the bathroom.

\subsection{Lessons from operating across studies}

Three patterns recur across the use cases.

\textit{Hotels accept sensors that need no attention.} Guests can see most of our sensors, and that has rarely been the problem. The temperature and humidity sensors stick to the wall with removable tape, run for over a year on a battery, and nobody touches them again. The smart plugs for the minifridges sit between the fridge cable and the wall socket. The people counters stick to the ceiling. None of this creates work for housekeeping or maintenance. The difficult sensors are the ones that do. The food waste scales need charging every week and resetting to zero before each service. The bin fill-level sensors are mounted on the ceiling rather than the bin lid, because that is what hotels prefer, so staff have to return the bin to the same spot every time they empty it. The power clamps have to go around the phase conductor inside the mains wiring, which means a licensed electrician and a live circuit. When a hotel will not take on this work, we measure something else instead and estimate the answer from it. Humidity tells us that a guest has showered. Room temperature tells us the air conditioner is running. These sensors install in minutes and need nothing afterwards, but an estimate is not a measurement. LoRIS supports both kinds, and several studies use both. Which one we choose depends on what the hotel will maintain, how the room is built, and how accurate the study needs to be.

\textit{How often a sensor reports depends on how fast the behaviour is.} The EM300-TH sensors report every minute. Most LoRaWAN deployments report far less often. We need this speed because the events we study are short. A shower raises the humidity for 5--15 minutes. An air conditioner changes the room temperature over 10--30 minutes. If we sampled every half hour, these events would disappear into the surrounding readings and we would either miss them or get them wrong. Not every study needs this. Bins fill up slowly, so those sensors report every ten minutes. We set the platform default to suit the hardest task.

\textit{An intervention can save nothing and still annoy guests.} The table-sign study did not reduce plate waste \cite{zinn_not_2026}. Two of the four messages also slightly reduced guest satisfaction with their meal. Hotels will not use an intervention that upsets guests, however good it is for the environment. We only learned this because, in addition to the plate waste measured by our LoRIS system, we also measured guest satisfaction.

\section{Conclusions}
\label{sec:conclusions}

We present LoRIS, a large-scale LoRaWAN platform for sustainability analytics in hotels. LoRIS instruments 21 sites across Australia and Slovenia with 850 sensors of 19 types, since February 2022. It has collected over 202 million records across the AU915 and EU868 regulatory regions. The platform follows the standard LoRaWAN reference model and runs on managed cloud services. A serverless pipeline decodes, enriches, and stores each uplink, and grows with the number of sensors without capacity planning. Hotel managers, sustainability officers, and researchers read the data through role-based dashboards. We built the platform to fit the way hotels work and the privacy they owe their guests. Gateways open outbound connections only and expose no service to the hotel network. Encryption covers the whole path from sensor to dashboard. We chose the sensor types and decided where to place them so that the dataset holds no personally identifiable information in the first place.

LoRIS has supported seven field studies across food waste, energy consumption, and water consumption. Each study uses a different combination of sensors, reporting intervals, and inference methods on the same infrastructure. Three lessons recur. Hotels accept sensors that need no attention after installation, so the sensors that ask the least of staff are often the ones we can deploy. How often a sensor reports has to match how fast the behaviour is, which is why the EM300-TH sensors report every minute. Guest satisfaction needs measuring alongside the intervention, because an intervention that upsets guests will not be adopted however well it performs environmentally. To our knowledge, the resulting dataset is one of the largest LoRaWAN datasets collected for hotel sustainability research, in both record count and deployment duration.

We plan to release the dataset to the wider research community, pending ethics approval. As new partner agreements expand the deployment, we expect LoRIS to support a wider range of operational and behavioural studies, and to serve as a template for sustainability sensing in other hospitality and accommodation settings.

\section*{CRediT authorship contribution statement}
\textbf{Yash Pandey:} Conceptualization, Investigation, Software, Data curation, Formal analysis, Methodology, Visualization, Writing – original draft, Writing – review \& editing.
\textbf{Angus Gray:} Conceptualization, Software, Investigation.
\textbf{Reza Serati:} Investigation, Data curation, Formal analysis, Visualization, Writing – original draft, Writing – review \& editing.
\textbf{Oscar Zhu:} Investigation, Project administration, Resources.
\textbf{Emil Juvan:} Investigation, Resources.
\textbf{Anna Zinn:} Investigation, Resources.
\textbf{Danyelle Greene:} Investigation, Resources.
\textbf{Qingqing Chen:} Investigation, Resources.
\textbf{Sarah MacInnes:} Investigation, Resources.
\textbf{Siamak Layeghy:} Conceptualization, Methodology, Supervision, Writing – original draft, Writing – review \& editing.
\textbf{Sara Dolnicar:} Conceptualization, Methodology, Funding acquisition, Project administration, Resources, Supervision, Writing – original draft, Writing – review \& editing.
\textbf{Marius Portmann:} Conceptualization, Methodology, Project administration, Resources, Supervision, Writing – original draft, Writing – review \& editing.

\section*{Acknowledgements}

This research was supported by the Australian Research Council through the Laureate Fellowship scheme (project number FL190100143), the Discovery Project scheme (project number DP250101460), and the Linkage Project scheme (project number LP200301583). The authors thank the 21 accommodation providers in Australia and Slovenia that hosted the deployment. Their managers and housekeeping teams accommodated sensor installation, maintenance, and battery replacement over multiple years, and several provided operational data and permission to include images taken on site. Individual properties are not named at their request.







\bibliography{references}

@inproceedings{bor_lora_2016,
	address = {New York, NY, USA},
	series = {{MSWiM} '16},
	title = {Do {LoRa} {Low}-{Power} {Wide}-{Area} {Networks} {Scale}?},
	isbn = {978-1-4503-4502-6},
	doi = {10.1145/2988287.2989163},
	urldate = {2026-03-16},
	booktitle = {Proceedings of the 19th {ACM} {International} {Conference} on {Modeling}, {Analysis} and {Simulation} of {Wireless} and {Mobile} {Systems}},
	publisher = {Association for Computing Machinery},
	author = {Bor, Martin C. and Roedig, Utz and Voigt, Thiemo and Alonso, Juan M.},
	month = nov,
	year = {2016},
	pages = {59--67},
}

@article{moon_hotel_2022,
	title = {Hotel privacy management and guest trust building: {A} relational signaling perspective},
	volume = {102},
	issn = {0278-4319},
	shorttitle = {Hotel privacy management and guest trust building},
	doi = {10.1016/j.ijhm.2022.103171},
	urldate = {2026-03-17},
	journal = {International Journal of Hospitality Management},
	author = {Moon, Hyoungeun and Yu, Jongsik and Chua, Bee-Lia and Han, Heesup},
	month = apr,
	year = {2022},
	pages = {103171},
}

@article{serati_deep_2026,
	title = {Deep sequence learning models for hotel room air conditioner runtime estimation via ambient sensing},
	issn = {0378-7788},
	doi = {10.1016/j.enbuild.2026.118118},
	urldate = {2026-08-17},
	journal = {Energy and Buildings},
	author = {Serati, Reza and Kulatilleke, Gayan and Layeghy, Siamak and Ariai, Farid and Dolnicar, Sara and Portmann, Marius},
	month = aug,
	year = {2026},
	pages = {118118},
}

@article{frei_building_nodate,
	title = {Building {Energy} {Performance} {Assessment} {Using} an {Easily} {Deployable} {Sensor} {Kit}: {Process}, {Risks}, and {Lessons} {Learned}},
	volume = {6},
	issn = {2297-3362},
	shorttitle = {Building {Energy} {Performance} {Assessment} {Using} an {Easily} {Deployable} {Sensor} {Kit}},
	doi = {10.3389/fbuil.2020.609877},
	language = {English},
	urldate = {2026-03-17},
	journal = {Frontiers in Built Environment},
	publisher = {Frontiers},
	author = {Frei, Mario and Deb, Chirag and Nagy, Zoltan and Hischier, Illias and Schlueter, Arno},
}

@article{dolnicar_automatically_2023,
	title = {Automatically monitoring environmental performance in tourism – {The} example of plate waste at all-you-can-eat buffets},
	volume = {4},
	issn = {2666-9579},
	doi = {10.1016/j.annale.2023.100100},
	number = {2},
	urldate = {2026-04-09},
	journal = {Annals of Tourism Research Empirical Insights},
	author = {Dolnicar, Sara and Gray, Angus and Grün, Bettina and Li, Hongwei and Portmann, Marius},
	month = nov,
	year = {2023},
	pages = {100100},
}

@inproceedings{lavdas_performance_2025,
	title = {Performance {Evaluation} of {LoRaWAN} {Networks} for {Smart} {Water} {Metering}},
	issn = {2325-2944},
	doi = {10.1109/DCOSS-IoT65416.2025.00141},
	urldate = {2026-06-28},
	booktitle = {2025 21st {International} {Conference} on {Distributed} {Computing} in {Smart} {Systems} and the {Internet} of {Things} ({DCOSS}-{IoT})},
	author = {Lavdas, Spyros and Vardoulias, George and El Hajj, Wassim and Zinonos, Zinon},
	month = jun,
	year = {2025},
	note = {ISSN: 2325-2944},
	pages = {929--935},
}

@article{sablica_ecoshower_2025,
	title = {{EcoShower}: {Estimating} shower duration using non-intrusive multi-modal sensor data via {LSTM} and {Gated} {Transformer} models},
	volume = {277},
	issn = {0957-4174},
	shorttitle = {{EcoShower}},
	doi = {10.1016/j.eswa.2025.127202},
	urldate = {2026-04-16},
	journal = {Expert Systems with Applications},
	author = {Sablica, Lukas and Grün, Bettina and Layeghy, Siamak and Dolnicar, Sara and Portmann, Marius},
	month = jun,
	year = {2025},
	pages = {127202},
}

@article{dolnicar_does_2024,
	title = {Does every hotel room need a minifridge? {Empirical} evidence from consumer self-reports and an automatic sensor-based system measuring electricity consumption and guest use},
	volume = {5},
	issn = {2666-9579},
	shorttitle = {Does every hotel room need a minifridge?},
	doi = {10.1016/j.annale.2024.100134},
	number = {2},
	urldate = {2026-04-09},
	journal = {Annals of Tourism Research Empirical Insights},
	author = {Dolnicar, Sara and Greene, Danyelle and Layeghy, Siamak and Portmann, Marius},
	month = nov,
	year = {2024},
	pages = {100134},
}

@article{greene_crikey_2025,
	title = {“{Crikey}! {Let}'s keep it cozy like a joey in a pouch” can humour or compassion encourage sustainable heater use at hotels?},
	volume = {107},
	issn = {0272-4944},
	doi = {10.1016/j.jenvp.2025.102779},
	urldate = {2026-04-16},
	journal = {Journal of Environmental Psychology},
	author = {Greene, Danyelle and Zinn, Anna K. and Chen, Qingqing and Serati, Reza and Portmann, Marius and Dolnicar, Sara},
	month = nov,
	year = {2025},
	pages = {102779},
}

@article{greene_leveraging_2025,
	title = {Leveraging social norms and empathy to encourage sustainable air conditioning practices amongst hotel guests},
	issn = {0272-4944},
	doi = {10.1016/j.jenvp.2025.102811},
	urldate = {2025-10-20},
	journal = {Journal of Environmental Psychology},
	author = {Greene, Danyelle and Birenboim, Amit and Zinn, Anna K. and Portmann, Marius and Pandey, Yash and Grün, Bettina and Dolnicar, Sara},
	month = oct,
	year = {2025},
	pages = {102811},
}

@article{teymuri_lp-mab_2023,
	title = {{LP}-{MAB}: {Improving} the {Energy} {Efficiency} of {LoRaWAN} {Using} a {Reinforcement}-{Learning}-{Based} {Adaptive} {Configuration} {Algorithm}},
	volume = {23},
	copyright = {http://creativecommons.org/licenses/by/3.0/},
	issn = {1424-8220},
	shorttitle = {{LP}-{MAB}},
	doi = {10.3390/s23042363},
	language = {en},
	number = {4},
	urldate = {2026-07-05},
	journal = {Sensors},
	publisher = {Multidisciplinary Digital Publishing Institute},
	author = {Teymuri, Benyamin and Serati, Reza and Anagnostopoulos, Nikolaos Athanasios and Rasti, Mehdi},
	month = jan,
	year = {2023},
	pages = {2363},
}

@article{rosa_development_2026,
	title = {Development and validation of an integrated {IoT} system for monitoring barn environment, gaseous concentrations and slurry management in dairy cattle farms},
	volume = {37},
	issn = {2542-6605},
	doi = {10.1016/j.iot.2026.101914},
	urldate = {2026-05-03},
	journal = {Internet of Things},
	author = {Rosa, E. and Rincón, L. and Merino, P.},
	month = may,
	year = {2026},
	pages = {101914},
}

@article{chan_iot_2023,
	title = {{IoT} devices deployment challenges and studies in building management system},
	volume = {2},
	issn = {2813-3110},
	doi = {10.3389/friot.2023.1254160},
	language = {English},
	urldate = {2026-03-17},
	journal = {Frontiers in the Internet of Things},
	publisher = {Frontiers},
	author = {Chan, Raymond and Yan, Wye Kaye and Ma, Jung Man and Loh, Kai Mun and Yu, Tan and Low, Malcolm Yoke Hean and Yar, Kar Peo and Rehman, Habib and Phua, Thong Chee},
	month = dec,
	year = {2023},
}

@article{sakariyah_adewole_systematic_2025,
	title = {A {Systematic} {Literature} {Review} of {Privacy} {Related} to {Sensing} in {Smart} {Buildings}},
	volume = {13},
	issn = {2169-3536},
	doi = {10.1109/ACCESS.2025.3611344},
	urldate = {2026-03-17},
	journal = {IEEE Access},
	author = {Sakariyah Adewole, Kayode and Persson, Jan A. and Jacobsson, Andreas and Akin, Erdal and Shokrollahi, Azad and Malekian, Reza and Caltenco, Héctor and Valtonen Örnhag, Marcus},
	year = {2025},
	pages = {164358--164394},
}

@article{karadayi-usta_cybersecurity_2024,
	title = {Cybersecurity {Risks} {Analysis} in the {Hospitality} {Industry}: {A} {Stakeholder} {Perspective} on {Sustainable} {Service} {Systems}},
	volume = {12},
	copyright = {http://creativecommons.org/licenses/by/3.0/},
	issn = {2079-8954},
	shorttitle = {Cybersecurity {Risks} {Analysis} in the {Hospitality} {Industry}},
	doi = {10.3390/systems12100397},
	language = {en},
	number = {10},
	urldate = {2026-03-17},
	journal = {Systems},
	publisher = {Multidisciplinary Digital Publishing Institute},
	author = {Karadayi-Usta, Saliha},
	month = oct,
	year = {2024},
	pages = {397},
}

@article{wynn_it_2022,
	title = {{IT} {Strategy} in the {Hotel} {Industry} in the {Digital} {Era}},
	volume = {14},
	copyright = {http://creativecommons.org/licenses/by/3.0/},
	issn = {2071-1050},
	doi = {10.3390/su141710705},
	language = {en},
	number = {17},
	urldate = {2026-03-17},
	journal = {Sustainability},
	publisher = {Multidisciplinary Digital Publishing Institute},
	author = {Wynn, Martin and Jones, Peter},
	month = jan,
	year = {2022},
	pages = {10705},
}

@article{harinda_performance_2022,
	title = {Performance of a {Live} {Multi}-{Gateway} {LoRaWAN} and {Interference} {Measurement} across {Indoor} and {Outdoor} {Localities}},
	volume = {11},
	copyright = {http://creativecommons.org/licenses/by/3.0/},
	issn = {2073-431X},
	doi = {10.3390/computers11020025},
	language = {en},
	number = {2},
	urldate = {2026-06-28},
	journal = {Computers},
	publisher = {Multidisciplinary Digital Publishing Institute},
	author = {Harinda, Eugen and Wixted, Andrew J. and Qureshi, Ayyaz-UI-Haq and Larijani, Hadi and Gibson, Ryan M.},
	month = feb,
	year = {2022},
	pages = {25},
}

@article{basford_lorawan_2020,
	title = {{LoRaWAN} for {Smart} {City} {IoT} {Deployments}: {A} {Long} {Term} {Evaluation}},
	volume = {20},
	copyright = {http://creativecommons.org/licenses/by/3.0/},
	issn = {1424-8220},
	shorttitle = {{LoRaWAN} for {Smart} {City} {IoT} {Deployments}},
	doi = {10.3390/s20030648},
	language = {en},
	number = {3},
	urldate = {2026-02-18},
	journal = {Sensors},
	publisher = {Multidisciplinary Digital Publishing Institute},
	author = {Basford, Philip J. and Bulot, Florentin M. J. and Apetroaie-Cristea, Mihaela and Cox, Simon J. and Ossont, Steven J.},
	month = jan,
	year = {2020},
	pages = {648},
}

@article{yasmin_lorawan_2020,
	title = {{LoRaWAN} for {Smart} {Campus}: {Deployment} and {Long}-{Term} {Operation} {Analysis}},
	volume = {20},
	copyright = {http://creativecommons.org/licenses/by/3.0/},
	issn = {1424-8220},
	shorttitle = {{LoRaWAN} for {Smart} {Campus}},
	doi = {10.3390/s20236721},
	language = {en},
	number = {23},
	urldate = {2026-05-03},
	journal = {Sensors},
	publisher = {Multidisciplinary Digital Publishing Institute},
	author = {Yasmin, Rumana and Mikhaylov, Konstantin and Pouttu, Ari},
	month = jan,
	year = {2020},
	pages = {6721},
}

@article{gaffurini_end--end_2024,
	title = {End-to-{End} {Emulation} of {LoRaWAN} {Architecture} and {Infrastructure} in {Complex} {Smart} {City} {Scenarios} {Exploiting} {Containers}},
	volume = {24},
	copyright = {http://creativecommons.org/licenses/by/3.0/},
	issn = {1424-8220},
	doi = {10.3390/s24072024},
	language = {en},
	number = {7},
	urldate = {2026-06-28},
	journal = {Sensors},
	publisher = {Multidisciplinary Digital Publishing Institute},
	author = {Gaffurini, Massimiliano and Flammini, Alessandra and Ferrari, Paolo and Fernandes Carvalho, Dhiego and Godoy, Eduardo Paciencia and Sisinni, Emiliano},
	month = jan,
	year = {2024},
	pages = {2024},
}

@article{lavdas_evaluating_2025,
	title = {Evaluating {LoRaWAN} {Network} {Performance} in {Smart} {City} {Environments} {Using} {Machine} {Learning}},
	volume = {12},
	issn = {2327-4662},
	doi = {10.1109/JIOT.2025.3562222},
	number = {14},
	urldate = {2026-06-28},
	journal = {IEEE Internet of Things Journal},
	author = {Lavdas, Spyros and Bakas, Nikolaos and Vavousis, Konstantinos and Khalifeh, Ala' and El Hajj, Wassim and Zinonos, Zinon},
	month = jul,
	year = {2025},
	pages = {27060--27074},
}

@article{kufakunesu_survey_2020,
	title = {A {Survey} on {Adaptive} {Data} {Rate} {Optimization} in {LoRaWAN}: {Recent} {Solutions} and {Major} {Challenges}},
	volume = {20},
	copyright = {http://creativecommons.org/licenses/by/3.0/},
	issn = {1424-8220},
	shorttitle = {A {Survey} on {Adaptive} {Data} {Rate} {Optimization} in {LoRaWAN}},
	doi = {10.3390/s20185044},
	language = {en},
	number = {18},
	urldate = {2026-03-17},
	journal = {Sensors},
	publisher = {Multidisciplinary Digital Publishing Institute},
	author = {Kufakunesu, Rachel and Hancke, Gerhard P. and Abu-Mahfouz, Adnan M.},
	month = jan,
	year = {2020},
	pages = {5044},
}

@article{casals_modeling_2017,
	title = {Modeling the {Energy} {Performance} of {LoRaWAN}},
	volume = {17},
	copyright = {http://creativecommons.org/licenses/by/3.0/},
	issn = {1424-8220},
	doi = {10.3390/s17102364},
	language = {en},
	number = {10},
	urldate = {2026-03-17},
	journal = {Sensors},
	publisher = {Multidisciplinary Digital Publishing Institute},
	author = {Casals, Lluís and Mir, Bernat and Vidal, Rafael and Gomez, Carles},
	month = oct,
	year = {2017},
	pages = {2364},
}

@article{jouhari_survey_2023,
	title = {A {Survey} on {Scalable} {LoRaWAN} for {Massive} {IoT}: {Recent} {Advances}, {Potentials}, and {Challenges}},
	volume = {25},
	issn = {1553-877X},
	shorttitle = {A {Survey} on {Scalable} {LoRaWAN} for {Massive} {IoT}},
	doi = {10.1109/COMST.2023.3274934},
	number = {3},
	urldate = {2026-05-04},
	journal = {IEEE Communications Surveys \& Tutorials},
	author = {Jouhari, Mohammed and Saeed, Nasir and Alouini, Mohamed-Slim and Amhoud, El Mehdi},
	year = {2023},
	pages = {1841--1876},
}

@article{shanmuga_sundaram_survey_2020,
	title = {A {Survey} on {LoRa} {Networking}: {Research} {Problems}, {Current} {Solutions}, and {Open} {Issues}},
	volume = {22},
	issn = {1553-877X},
	shorttitle = {A {Survey} on {LoRa} {Networking}},
	doi = {10.1109/COMST.2019.2949598},
	number = {1},
	urldate = {2026-06-28},
	journal = {IEEE Communications Surveys \& Tutorials},
	author = {Shanmuga Sundaram, Jothi Prasanna and Du, Wan and Zhao, Zhiwei},
	month = jan,
	year = {2020},
	pages = {371--388},
}

@article{adelantado_understanding_2017,
	title = {Understanding the {Limits} of {LoRaWAN}},
	volume = {55},
	issn = {1558-1896},
	doi = {10.1109/MCOM.2017.1600613},
	number = {9},
	urldate = {2026-03-17},
	journal = {IEEE Communications Magazine},
	author = {Adelantado, Ferran and Vilajosana, Xavier and Tuset-Peiro, Pere and Martinez, Borja and Melia-Segui, Joan and Watteyne, Thomas},
	month = sep,
	year = {2017},
	pages = {34--40},
}

@inproceedings{tong_citywide_2024,
	address = {New York, NY, USA},
	series = {{SenSys} '23},
	title = {Citywide {LoRa} {Network} {Deployment} and {Operation}: {Measurements}, {Analysis}, and {Implications}},
	isbn = {979-8-4007-0414-7},
	shorttitle = {Citywide {LoRa} {Network} {Deployment} and {Operation}},
	doi = {10.1145/3625687.3625796},
	urldate = {2026-06-23},
	booktitle = {Proceedings of the 21st {ACM} {Conference} on {Embedded} {Networked} {Sensor} {Systems}},
	publisher = {Association for Computing Machinery},
	author = {Tong, Shuai and Wang, Jiliang and Yang, Jing and Liu, Yunhao and Zhang, Jun},
	month = apr,
	year = {2024},
	pages = {362--375},
}

@article{georgiou_low_2017,
	title = {Low {Power} {Wide} {Area} {Network} {Analysis}: {Can} {LoRa} {Scale}?},
	volume = {6},
	issn = {2162-2345},
	shorttitle = {Low {Power} {Wide} {Area} {Network} {Analysis}},
	doi = {10.1109/LWC.2016.2647247},
	number = {2},
	urldate = {2026-06-28},
	journal = {IEEE Wireless Communications Letters},
	author = {Georgiou, Orestis and Raza, Usman},
	month = apr,
	year = {2017},
	pages = {162--165},
}

@article{haxhibeqiri_survey_2018,
	title = {A {Survey} of {LoRaWAN} for {IoT}: {From} {Technology} to {Application}},
	volume = {18},
	copyright = {http://creativecommons.org/licenses/by/3.0/},
	issn = {1424-8220},
	shorttitle = {A {Survey} of {LoRaWAN} for {IoT}},
	doi = {10.3390/s18113995},
	language = {en},
	number = {11},
	urldate = {2026-07-02},
	journal = {Sensors},
	publisher = {Multidisciplinary Digital Publishing Institute},
	author = {Haxhibeqiri, Jetmir and De Poorter, Eli and Moerman, Ingrid and Hoebeke, Jeroen},
	month = nov,
	year = {2018},
	pages = {3995},
}

@article{mekki_comparative_2019,
	title = {A comparative study of {LPWAN} technologies for large-scale {IoT} deployment},
	volume = {5},
	issn = {2405-9595},
	doi = {10.1016/j.icte.2017.12.005},
	number = {1},
	urldate = {2026-03-17},
	journal = {ICT Express},
	author = {Mekki, Kais and Bajic, Eddy and Chaxel, Frederic and Meyer, Fernand},
	month = mar,
	year = {2019},
	pages = {1--7},
}

@article{raza_low_2017,
	title = {Low {Power} {Wide} {Area} {Networks}: {An} {Overview}},
	volume = {19},
	issn = {1553-877X},
	shorttitle = {Low {Power} {Wide} {Area} {Networks}},
	doi = {10.1109/COMST.2017.2652320},
	number = {2},
	urldate = {2026-06-28},
	journal = {IEEE Communications Surveys \& Tutorials},
	author = {Raza, Usman and Kulkarni, Parag and Sooriyabandara, Mahesh},
	year = {2017},
	pages = {855--873},
}

@article{hassan_shmis_2025,
	title = {{SHMIS}: {An} integrated {IoT} context awareness framework for hotel management to enhance guest experience and operational efficiency},
	volume = {27},
	issn = {1943-4294},
	shorttitle = {{SHMIS}},
	doi = {10.1007/s40558-025-00316-4},
	language = {en},
	number = {3},
	urldate = {2026-06-28},
	journal = {Information Technology \& Tourism},
	author = {Hassan, Samah A. Z. and Eassa, Ahmed M.},
	month = sep,
	year = {2025},
	pages = {579--612},
}

@inproceedings{kansakar_fog-assisted_2018,
	title = {A {Fog}-{Assisted} {Architecture} to {Support} an {Evolving} {Hospitality} {Industry} in {Smart} {Cities}},
	issn = {2334-3141},
	doi = {10.1109/FIT.2018.00018},
	urldate = {2026-06-28},
	booktitle = {2018 {International} {Conference} on {Frontiers} of {Information} {Technology} ({FIT})},
	author = {Kansakar, Prasanna and Munir, Arslan and Shabani, Neda},
	month = dec,
	year = {2018},
	note = {ISSN: 2334-3141},
	pages = {59--64},
}

@article{kormos_validity_2014,
	title = {The validity of self-report measures of proenvironmental behavior: {A} meta-analytic review},
	volume = {40},
	issn = {02724944},
	shorttitle = {The validity of self-report measures of proenvironmental behavior},
	doi = {10.1016/j.jenvp.2014.09.003},
	language = {en},
	urldate = {2026-07-02},
	journal = {Journal of Environmental Psychology},
	author = {Kormos, Christine and Gifford, Robert},
	month = dec,
	year = {2014},
	pages = {359--371},
}

@article{bashir_reference_2022,
	title = {A {Reference} {Architecture} for {IoT}-{Enabled} {Smart} {Buildings}},
	volume = {3},
	issn = {2661-8907},
	doi = {10.1007/s42979-022-01401-9},
	language = {en},
	number = {6},
	urldate = {2026-06-28},
	journal = {SN Computer Science},
	author = {Bashir, Muhammad Rizwan and Gill, Asif Qumer and Beydoun, Ghassan},
	month = sep,
	year = {2022},
	pages = {493},
}

@article{al-fuqaha_internet_2015,
	title = {Internet of {Things}: {A} {Survey} on {Enabling} {Technologies}, {Protocols}, and {Applications}},
	volume = {17},
	issn = {1553-877X},
	shorttitle = {Internet of {Things}},
	doi = {10.1109/COMST.2015.2444095},
	number = {4},
	urldate = {2026-06-28},
	journal = {IEEE Communications Surveys \& Tutorials},
	author = {Al-Fuqaha, Ala and Guizani, Mohsen and Mohammadi, Mehdi and Aledhari, Mohammed and Ayyash, Moussa},
	month = oct,
	year = {2015},
	pages = {2347--2376},
}

@article{gubbi_internet_2013,
	title = {Internet of {Things} ({IoT}): {A} vision, architectural elements, and future directions},
	volume = {29},
	issn = {0167-739X},
	shorttitle = {Internet of {Things} ({IoT})},
	doi = {10.1016/j.future.2013.01.010},
	number = {7},
	urldate = {2026-06-28},
	journal = {Future Generation Computer Systems},
	author = {Gubbi, Jayavardhana and Buyya, Rajkumar and Marusic, Slaven and Palaniswami, Marimuthu},
	month = sep,
	year = {2013},
	pages = {1645--1660},
}

@article{guix_changing_2025,
	title = {The changing institutional logics behind sustainability reports from the largest hotel groups in the world in 2014, 2018 and 2021},
	volume = {106},
	issn = {0261-5177},
	doi = {10.1016/j.tourman.2024.105031},
	urldate = {2026-03-17},
	journal = {Tourism Management},
	author = {Guix, Mireia and Nájera Sánchez, Juan José and Bonilla Priego, M Jesús and Font, Xavier},
	month = feb,
	year = {2025},
	pages = {105031},
}

@article{juvan_attitudebehaviour_2014,
	title = {The attitude–behaviour gap in sustainable tourism},
	volume = {48},
	issn = {01607383},
	doi = {10.1016/j.annals.2014.05.012},
	language = {en},
	urldate = {2026-07-02},
	journal = {Annals of Tourism Research},
	author = {Juvan, Emil and Dolnicar, Sara},
	month = sep,
	year = {2014},
	pages = {76--95},
}

@article{juvan_measuring_2016,
	title = {Measuring environmentally sustainable tourist behaviour},
	volume = {59},
	issn = {0160-7383},
	doi = {10.1016/j.annals.2016.03.006},
	urldate = {2026-08-11},
	journal = {Annals of Tourism Research},
	author = {Juvan, Emil and Dolnicar, Sara},
	month = jul,
	year = {2016},
	pages = {30--44},
}

@article{zinn_not_2026,
	title = {Not worth the paper they are printed on? {The} effectiveness of table signs in reducing buffet plate waste},
	volume = {113},
	copyright = {All rights reserved},
	issn = {0261-5177},
	shorttitle = {Not worth the paper they are printed on?},
	doi = {10.1016/j.tourman.2025.105348},
	urldate = {2026-02-13},
	journal = {Tourism Management},
	author = {Zinn, Anna K. and Greene, Danyelle and Kozlov, Sergey and Grün, Bettina and Pandey, Yash and Portmann, Marius and Dolnicar, Sara},
	month = apr,
	year = {2026},
	pages = {105348},
}

@article{juvan_importance_2025,
	title = {On the {Importance} of {Field} {Studies} for {Testing} {Theory}-{Driven} {Behavioral} {Change} {Interventions} in ({Sustainable}) {Tourism}},
	volume = {64},
	issn = {0047-2875},
	doi = {10.1177/00472875241253009},
	language = {EN},
	number = {6},
	urldate = {2026-04-16},
	journal = {Journal of Travel Research},
	publisher = {SAGE Publications Inc},
	author = {Juvan, Emil and Zhu, Oscar Yuheng and Grün, Bettina and Dolnicar, Sara},
	month = jul,
	year = {2025},
	pages = {1449--1463},
}

@article{juvan_biting_2018,
	title = {Biting {Off} {More} {Than} {They} {Can} {Chew}: {Food} {Waste} at {Hotel} {Breakfast} {Buffets}},
	volume = {57},
	issn = {0047-2875},
	shorttitle = {Biting {Off} {More} {Than} {They} {Can} {Chew}},
	doi = {10.1177/0047287516688321},
	language = {EN},
	number = {2},
	urldate = {2026-08-13},
	journal = {Journal of Travel Research},
	publisher = {SAGE Publications Inc},
	author = {Juvan, Emil and Grün, Bettina and Dolnicar, Sara},
	month = feb,
	year = {2018},
	pages = {232--242},
}

@article{kasavan_drivers_2022,
	title = {Drivers of food waste generation and best practice towards sustainable food waste management in the hotel sector: a systematic review},
	volume = {29},
	issn = {1614-7499},
	shorttitle = {Drivers of food waste generation and best practice towards sustainable food waste management in the hotel sector},
	doi = {10.1007/s11356-022-19984-4},
	language = {en},
	number = {32},
	urldate = {2026-03-17},
	journal = {Environmental Science and Pollution Research},
	author = {Kasavan, Saraswathy and Siron, Rusinah and Yusoff, Sumiani and Fakri, Mohd Fadhli Rahmat},
	month = jul,
	year = {2022},
	pages = {48152--48167},
}

@article{juvan_drivers_2021,
	title = {Drivers of plate waste at buffets: {A} comprehensive conceptual model based on observational data and staff insights},
	volume = {2},
	issn = {2666-9579},
	shorttitle = {Drivers of plate waste at buffets},
	doi = {10.1016/j.annale.2021.100010},
	number = {1},
	urldate = {2026-04-16},
	journal = {Annals of Tourism Research Empirical Insights},
	author = {Juvan, Emil and Grün, Bettina and Zabukovec Baruca, Petra and Dolnicar, Sara},
	month = may,
	year = {2021},
	pages = {100010},
}

@article{becken_evidence_2017,
	title = {Evidence of the water-energy nexus in tourist accommodation},
	volume = {144},
	issn = {0959-6526},
	doi = {10.1016/j.jclepro.2016.12.167},
	urldate = {2026-03-17},
	journal = {Journal of Cleaner Production},
	author = {Becken, Susanne and McLennan, Char-lee},
	month = feb,
	year = {2017},
	pages = {415--425},
}

@article{torres_heating_2020,
	series = {Technologies and {Materials} for {Renewable} {Energy}, {Environment} and {Sustainability}},
	title = {Heating ventilation and air-conditioned configurations for hotelsan approach review for the design and exploitation},
	volume = {6},
	issn = {2352-4847},
	doi = {10.1016/j.egyr.2020.09.026},
	urldate = {2026-03-17},
	journal = {Energy Reports},
	author = {Torres, Yamile Díaz and Herrera, Hernán Hernández and Plasencia, Mario A. Alvares Guerra and Novo, Eduardo Pérez and Cabrera, Lester Pimentel and Haeseldonckx, Dries and Silva-Ortega, Jorge Iván},
	month = nov,
	year = {2020},
	pages = {487--497},
}

@article{bohdanowicz_determinants_2007,
	title = {Determinants and benchmarking of resource consumption in hotels—{Case} study of {Hilton} {International} and {Scandic} in {Europe}},
	volume = {39},
	issn = {0378-7788},
	doi = {10.1016/j.enbuild.2006.05.005},
	number = {1},
	urldate = {2026-03-17},
	journal = {Energy and Buildings},
	author = {Bohdanowicz, Paulina and Martinac, Ivo},
	month = jan,
	year = {2007},
	pages = {82--95},
}

@article{gossling_review_2023,
	title = {A review of tourism and climate change mitigation: {The} scales, scopes, stakeholders and strategies of carbon management},
	volume = {95},
	issn = {0261-5177},
	shorttitle = {A review of tourism and climate change mitigation},
	doi = {10.1016/j.tourman.2022.104681},
	urldate = {2026-03-17},
	journal = {Tourism Management},
	author = {Gössling, Stefan and Balas, Martin and Mayer, Marius and Sun, Ya-Yen},
	month = apr,
	year = {2023},
	pages = {104681},
}

@article{lenzen_carbon_2018,
	title = {The carbon footprint of global tourism},
	volume = {8},
	copyright = {2018 The Author(s)},
	issn = {1758-6798},
	doi = {10.1038/s41558-018-0141-x},
	language = {en},
	number = {6},
	journal = {Nature Climate Change},
	publisher = {Nature Publishing Group},
	author = {Lenzen, Manfred and Sun, Ya-Yen and Faturay, Futu and Ting, Yuan-Peng and Geschke, Arne and Malik, Arunima},
	month = jun,
	year = {2018},
	pages = {522--528},
}

@article{sun_drivers_2024,
	title = {Drivers of global tourism carbon emissions},
	volume = {15},
	copyright = {2024 The Author(s)},
	issn = {2041-1723},
	doi = {10.1038/s41467-024-54582-7},
	language = {en},
	number = {1},
	journal = {Nature Communications},
	publisher = {Nature Publishing Group},
	author = {Sun, Ya-Yen and Faturay, Futu and Lenzen, Manfred and Gössling, Stefan and Higham, James},
	month = dec,
	year = {2024},
	pages = {10384},
}

@misc{noauthor_kibana_nodate,
	title = {Kibana: {Visualize}, explore, and manage data in {Elasticsearch}},
	shorttitle = {Kibana},
	url = {https://www.elastic.co/kibana},
	language = {en},
	urldate = {2026-07-31},
	journal = {Elastic},
}

@misc{noauthor_auth0_nodate,
	title = {Auth0 {\textbar} {Why} {Auth0}?},
	url = {https://auth0.com/resources/ebooks/why-auth0-by-okta},
	language = {en},
	urldate = {2026-07-31},
	journal = {Auth0},
}

@misc{noauthor_elasticsearch_nodate,
	title = {Elasticsearch: {The} {Official} {Distributed} {Search} \& {Analytics} {Engine}},
	shorttitle = {Elasticsearch},
	url = {https://www.elastic.co/elasticsearch},
	language = {en},
	urldate = {2026-07-31},
	journal = {Elastic},
}

@article{pirani_solid_2014,
	title = {Solid waste management in the hospitality industry: {A} review},
	volume = {146},
	issn = {0301-4797},
	shorttitle = {Solid waste management in the hospitality industry},
	doi = {10.1016/j.jenvman.2014.07.038},
	journal = {Journal of Environmental Management},
	author = {Pirani, Sanaa I. and Arafat, Hassan A.},
	month = dec,
	year = {2014},
	pages = {320--336},
}

@article{atzori_internet_2010,
	title = {The {Internet} of {Things}: {A} survey},
	volume = {54},
	issn = {1389-1286},
	doi = {10.1016/J.COMNET.2010.05.010},
	number = {15},
	urldate = {2023-03-22},
	journal = {Computer Networks},
	publisher = {Elsevier},
	author = {Atzori, Luigi and Iera, Antonio and Morabito, Giacomo},
	month = oct,
	year = {2010},
	pages = {2787--2805},
}

@misc{noauthor_rp2-102_2020,
	title = {{RP2}-1.0.2 {LoRaWAN}® {Regional} {Parameters}},
	url = {https://resources.lora-alliance.org/technical-specifications/rp2-1-0-2-lorawan-regional-parameters},
	language = {en-US},
	urldate = {2026-05-03},
	journal = {LoRa Alliance},
	month = oct,
	year = {2020},
}

@misc{noauthor_ts001-104_2023,
	title = {{TS001}-1.0.4 {LoRaWAN}® {L2} 1.0.4 {Specification}},
	url = {https://resources.lora-alliance.org/technical-specifications/ts001-1-0-4-lorawan-l2-1-0-4-specification},
	language = {en-US},
	urldate = {2026-05-03},
	journal = {LoRa Alliance},
	month = sep,
	year = {2023},
}

@misc{environment_food_2024,
	title = {Food {Waste} {Index} {Report} 2024 {\textbar} {UNEP} - {UN} {Environment} {Programme}},
	url = {https://www.unep.org/resources/publication/food-waste-index-report-2024},
	language = {en},
	urldate = {2026-03-17},
	author = {Environment, U. N.},
	month = mar,
	year = {2024},
	note = {Section: publications},
}

@article{juvan_waste_2023,
	title = {Waste production patterns in hotels and restaurants: {An} intra-sectoral segmentation approach},
	volume = {4},
	issn = {2666-9579},
	shorttitle = {Waste production patterns in hotels and restaurants},
	doi = {10.1016/j.annale.2023.100090},
	number = {1},
	journal = {Annals of Tourism Research Empirical Insights},
	author = {Juvan, Emil and Grün, Bettina and Dolnicar, Sara},
	month = may,
	year = {2023},
	pages = {100090},
}

@article{dolnicar_designing_2020,
	title = {Designing for {More} {Environmentally} {Friendly} {Tourism}},
	volume = {84},
	issn = {01607383},
	doi = {10.1016/J.ANNALS.2020.102933},
	urldate = {2023-04-08},
	journal = {Annals of Tourism Research},
	publisher = {Elsevier Ltd},
	author = {Dolnicar, Sara},
	month = sep,
	year = {2020},
}

\end{document}